\documentclass[journal,onecolumn]{IEEEtran}   %运行1下
\usepackage{graphicx}
\usepackage{epstopdf}
\usepackage{amssymb}
\usepackage{amsfonts}
\usepackage{amsmath}
\usepackage{algorithm}
\usepackage{algorithmic}
\usepackage{subeqnarray}
\usepackage{cases}
\usepackage{bm}
\usepackage{subfigure,amsmath,amssymb,cite}
\usepackage{stfloats}

\usepackage{float}
\usepackage{cuted}
\UseRawInputEncoding

\ifCLASSINFOpdf
\else
\fi
\begin{document}

\title{Two Efficient Beamforming Methods for Hybrid IRS-aided AF Relay Wireless Networks}

\author{Xuehui~Wang,~Feng Shu,~Mengxing Huang,~Fuhui Zhou,~Riqing Chen,~Cunhua Pan,\\~Yongpeng Wu,~and~Jiangzhou~Wang,~\emph{Fellow},~\emph{IEEE}

\thanks{This work was supported in part by the National Natural Science Foundation of China (Nos.U22A2002, and 62071234), the Hainan Province Science and Technology Special Fund (ZDKJ2021022), the Scientific Research Fund Project of Hainan University under Grant KYQD(ZR)-21008, and the Collaborative Innovation Center of Information Technology, Hainan University (XTCX2022XXC07). This article was presented in part at the IEEE International Conference on Intelligent Communications and Computing, Nanchang, China, November 2023. \emph{(Corresponding authors: Feng Shu)}.}

\thanks{Xuehui~Wang and Mengxing Huang are  with the School of Information and Communication Engineering, Hainan University,~Haikou,~570228, China.}

\thanks{Feng Shu is with the School of Information and Communication Engineering and Collaborative Innovation Center of Information Technology, Hainan University, Haikou 570228, China, and also with the School of Electronic and Optical Engineering, Nanjing University of Science and Technology, Nanjing 210094, China. (e-mail: shufeng0101@163.com).}

\thanks{Fuhui Zhou is with the College of Electronic and Information Engineering, Nanjing University of Aeronautics and Astronautics, Nanjing 210000, China,
also with the Key Laboratory of Dynamic Cognitive System of Electromagnetic Spectrum Space, Nanjing University of Aeronautics and Astronautics, Nanjing 210000, China, and also with the Ministry of Industry and Information Technology, Nanjing 211106, China. (e-mail: zhoufuhui@ieee.org).}

\thanks{Riqing Chen is with the Digital Fujian Institute of Big Data for Agriculture, Fujian Agriculture and Forestry University, Fuzhou 350002, China. (e-mail: riqing.chen@fafu.edu.cn).}

\thanks{Cunhua Pan is with National Mobile Communications Research Laboratory, Southeast University, Nanjing 211111, China. (e-mail: cpan@seu.edu.cn).}

\thanks{Yongpeng Wu is with the Shanghai Key Laboratory of Navigation and Location Based Services, Shanghai Jiao Tong University, Minhang 200240, China. (e-mail: yongpeng.wu2016@gmail.com).}

\thanks{Jiangzhou Wang is with the School of Engineering, University of Kent, Canterbury CT2 7NT, U.K. (e-mail: j.z.wang@kent.ac.uk).}

%\thanks{Xiaohu You is with the LEADS, the National Mobile Communications Research Laboratory, and the Frontiers Science Center for Mobile Information Communication and Security, Southeast University, Nanjing 211189, China; and also with the Purple Mountain Laboratories, Nanjing 211100, China. (e-mail: xhyu@seu.edu.cn). }

}

\maketitle

\begin{abstract}

Due to the ``double fading'' effect caused by conventional passive intelligent reflecting surface (IRS), the signal via the reflection link is weak. To enhance the received signal, active elements with the ability to amplify the reflected signal are introduced to the passive IRS forming hybrid IRS.
In this paper, we propose a hybrid IRS-aided amplify-and-forward (AF) relay wireless network, where an optimization problem is formulated, which is subject to the constraints of transmit power budgets at the source/AF relay/hybrid IRS and that of unit modulus for passive IRS elements. By alternately designing the beamforming matrix at AF relay and the reflecting coefficient matrices at IRS, signal-to-noise ratio can be maximized.
To achieve high rate performance and extend the coverage range, a high-performance method based on semidefinite relaxation and fractional programming (HP-SDR-FP) algorithm is presented. Due to its extremely high complexity, a low-complexity method based on whitening filter, general power iterative and generalized Rayleigh-Ritz (WF-GPI-GRR) is proposed, which is different from HP-SDR-FP method. It is assumed that the amplifying coefficient of each active IRS element is equal, and the corresponding analytical solution of the amplifying coefficient can be obtained according to the transmit powers at AF relay and hybrid IRS. Simulation results show that the proposed two methods can greatly improve the rate performance compared to the existing networks, such as the passive IRS-aided AF relay and only AF relay network. In particular, a 50.0\% rate gain over the existing networks is approximately achieved in the high power budget region of hybrid IRS. Moreover, it is verified that the proposed HP-SDR-FP method perform better than WF-GPI-GRR method in terms of rate performance.

\end{abstract}


\section{Introduction}

With the rapid expansion of Internet-of-Things (IoT), the scale of smart devices in communication network grows exponentially \cite{2020LC}. At present, because of high hardware cost and energy consumption, it is difficult for the existing technologies \cite{2021CJN}, such as wireless network coding, millimeter wave (mmWave), coordinated multi-point and massive multiple-input multiple-out (MIMO), to meet the stringent requirements, such as autonomous, ultra-large-scale, highly dynamic and fully intelligent services \cite{2017VWSW}. Deploying relay nodes in the existing wireless networks can save the cost of base stations and realize the cooperation of multiple communication nodes, so that extended coverage and high reliability can be obtained \cite{2017PS}. However, the relay is an active device, which consumes much power to forward signal. Therefore, it is imperative to develop a future wireless network, which is innovative, efficient and resource saving.

Because of the advantages of programmability, easy deployment, low cost and low power consumption, intelligent reflecting surface (IRS) is becoming popular, which has gained much research attention from both academia and industry \cite{2020WQQ}. IRS is composed of a large number of passive electromagnetic units, which are dynamically controlled to reflect incident signal forming an intelligent wireless propagation environment in a software-defined manner \cite{2021LY}. From the perspective of electromagnetic theory, radiation pattern and physics nature of IRS unit, the free-space path loss models for IRS-assisted wireless communications were well introduced in \cite{2021TWK}. A IRS-aided dual-hop visible light communication (VLC)/radio frequency (RF) system was proposed in \cite{2020YL1} , where the performance analysis related to the outage probability and bit error rate (BER) were presented. Because of reconfigurability, IRS has been viewed as an enabling and potential technology to achieve performance enhancement, spectral and energy efficiency improvement. With more and more research on IRS, IRS has been widely applied to the following scenarios, physical layer security \cite{ 2022JXY, 2021TY, 2019SH }, simultaneous wireless information and power transfer (SWIPT) \cite{2020WQQ1, 2021SWP}, multicell MIMO communications \cite{2020PCH, 2022RA}, covert communications \cite{2022ZXB, 2022CX}, and wireless powered communication network (WPCN) \cite{2022WQQ, 2021CHQ, 2022SWP}. To maximize secrecy rate for IRS-assisted multi-antenna systems in \cite{2019SH}, where an efficient alternating algorithm was developed to jointly optimize the transmit covariance of the source and the phase shift matrix of the IRS. For multicell communication systems \cite{2020PCH}, IRS was deployed at the cell boundary. While a method of jointly optimizing the active precoding matrices at the base stations (BSs) and the phase shifts at the IRS was proposed to maximize the weighted sum rate of all users. A IRS-aided secure MIMO WPCN is considered in \cite{2022SWP}, by jointly optimizing the downlink (DL)/uplink (UL) time allocation, the energy transmit covariance matrix of hybrid access point (AP), the transmit beamforming matrix of users and the phase shifts of IRS, the maximum secrecy throughput of all users was achieved.

Given the benefits of IRS and relay, their combination networks have emerged, which strike a good balance among hardware cost, energy consumption, coverage and rate performance. Recently, there were some related research works on the combination of IRS and relay appeared, which proved the combination could well serve for the wireless communication network in terms of coverage extension \cite{2020YL2, 2021YIK}, energy efficiency \cite{2021MO}, spectral efficiency \cite{2021NTN} and rate performance \cite{2022WXH, 2020AZD}. The authors proposed an IRS-assisted dual-hop free space optical and radio frequency (FSO-RF) communication system with a decode-and-forward (DF) relaying protocol, and derived the exact closed-form expressions for the outage probability and bit error rate (BER) \cite{2020YL2}. Meanwhile, the simulation results verified that the combination can expand the coverage. An IRS-aided multi-antenna DF relay network was proposed in \cite{2022WXH}, where three methods, an alternately iterative structure, null-space projection plus maximum ratio combining (MRC) and IRS element selection plus MRC, were put forward to improve the rate performance. Obviously, the rate performance was improved by optimizing beamforming at relay and phase shifts at IRS. Compared with an IRS-assisted wireless network, the authors in \cite{2020AZD} demonstrated that the combination network could utilize fewer IRS elements to harvest the same rate.

However, the above existing research work focused on conventional passive IRS. In fact, the reflected signal via passive IRS is weak due to the existence of ``double fading'' effect. Under such circumstances, the active IRS has emerged, which can amplify the reflected signal with extra power. In \cite{2022ZZJ}, the authors proposed the concept of active IRS and came up with a joint transmit and reflect precoding algorithm to solve the problem of capacity maximization, which existed in a signal model for active IRS. It was showed that the proposed active IRS could achieve a noticeable capacity gain compared to the existing passive IRS, which directly verified that active IRS could eliminate the ``double fading'' effect. With the same overall power budget, a fair performance comparison between active IRS and passive IRS was made theoretically in \cite{2022ZKD}, where it was proved that the active IRS is better than passive IRS in the case of a small or medium number of IRS elements or enough power budget. Accordingly, a novel active IRS-assisted secure wireless transmission was proposed in \cite{2022DLM}, where the non-convex secrecy rate optimization problem was solved by jointly optimizing the beamformer at transmitter and reflecting coefficient matrix at IRS. It was demonstrated that with the aid of active IRS, a significantly higher secrecy performance gain could be obtained compared with existing solutions with passive IRS and without IRS design.

Considering that active IRS has the ability to amplify signal, and aiming to further improve the rate performance or save more passive IRS elements of the combination network of passive IRS and relay, we propose that adding active IRS elements to passive IRS, thereby a hybrid
IRS-aided AF relay wireless network is generated, which makes full use of the advantages of passive IRS, active IRS and relay to strike a good balance among circuit cost, energy efficiency, coverage and rate performance. To our best knowledge, it is lack of little research work on the hybrid IRS-aided amplify-and-forward (AF) relay network, therefore, which motivates us to pay much attention to its further research.

%Therefore, it motivates us to pay much attention to the research of hybrid RIS-aided AF relay network.

In this case, using the criterion of Max SNR, two efficient beamforming methods are proposed to improve the rate performance of the proposed hybrid IRS-aided AF relay network or dramatically extend its coverage range. The main contributions of the paper are summarized as follows:

\begin{enumerate}

\item

%To make a dramatic rate improvement, a hybrid IRS-aided AF relay network model is proposed, where the hybrid IRS consists of passive IRS and active IRS.
To achieve a high rate, a high-performance method based on semidefinite relaxation and fractional programming (HP-SDR-FP) algorithm  is presented to jointly optimize the beamforming matrix at AF relay and the reflecting coefficient matrices at IRS by optimizing one and fixing the other two. However, it is difficult to directly solve the non-convex optimization problem with fractional and non-concave objective function and non-convex constraints.
To address this issue, some operations such as vectorization, Kronecker product and Hadamard product are applied to simplify the non-convex optimization problem, then SDR algorithm, Charnes-Cooper transformation of FP algorithm and Gaussian randomization method are adopted to obtain the optimization variable. The proposed HP-SDR-FP method can harvest up to 80\% rate gain over the passive IRS-aided AF relay network as the number of active IRS elements tends to large. Additionally, its convergence rate is fast, and its highest order of computational complexity is $M^{13}$ and $N^{6.5}$ FLOPs.

%The total computational complexity is extremely high, and the highest order of computational complexity is $M^{11}$ and $N^{5.5}$ FLOPs.

\item
%From the simulation results, the proposed Max-RP based on IRSES plus MRC method achieves a substantial rate improvement over existing system in \cite{2020A}. The complexity is $\mathcal{O}\{15MK+8M+10K+L_5(18MN+2M+3N)\}$, which is extremely lower than the above proposed two methods.

To further reduce the computational complexity of the above two methods, a low-complexity method based on whitening filter, general power iterative algorithm and generalized Rayleigh-Ritz theorem (WF-GPI-GRR) is put forward, where it is assumed that the amplifying coefficient of each active IRS element is equal in the first time slot or the second time slot. To exploit the colored property of noise, whitening filter operation is performed to the received signal. In line with the transmit power at AF relay and hybrid IRS, the analytical solution of the amplifying coefficient can be obtained. Moreover, the closed-form expression of beamforming matrix at AF relay is derived by utilizing maximum-ratio combining and maximum-ratio transmission (MRC-MRT) scheme, GPI and GRR are respectively applied to obtain the phase shift matrices at IRS for the first time slot and the second time slot. Compared with passive IRS-aided AF relay network, its rate can be improved by 49\%. Its highest order of computational complexity is $M^3$ and $N^3$ FLOPs, which is much lower than that of the proposed HP-SDR-FP method.

\end{enumerate}

The remainder of this paper is organized as follows. In Section \ref{SM}, a hybrid IRS-aided AF relay network is described, and an optimization problem is formulated. In Section \ref{SDR_FP}, we propose a high-performance method. A low-complexity method is presented in Section \ref{WF_GPI_GRR}. We present our simulation results in Section \ref{Results}, and draw conclusions in Section \ref{Conclusions}.

\emph{Notation}:
The letters of lower case, bold lower case and bold upper case are respectively applied to stand for scalars, vectors and matrices.
$(\cdot)^{-1}$, $(\cdot)^*$, $(\cdot)^T$ and $(\cdot)^H$ are adopted to represent the operations of inverse, matrix conjugate, transpose and conjugate transpose, respectively.
$\mathbb{E}\{\cdot\}$, $| \cdot |$, $\|\cdot\|$, $\text{tr}(\cdot)$ and $\text{arg}(\cdot)$ denote expectation operation, the modulus of a scalar, 2-norm, the trace of a matrix and the phase of a complex number, respectively.
$\odot$ and $\otimes$ respectively denote Hadamard and Kronecker product. $\textbf{I}_{N}$ is the $N$-dimensional identity matrix.

\section{System Model}\label{SM}
\subsection{Signal Model}
Fig. 1 sketches a hybrid IRS-aided AF relay network, where source (S) and destination (D) are respectively equipped with a single antenna, a AF relay working in half-duplex mode is with $M$ antennas, and a hybrid IRS is made up of $N$ units including $K$ active units and $L$ passive units. The active elements reflect the incident signal by adjusting the amplitude and phase, while the passive elements reflect the incident signal only by shifting the phase.
\begin{figure}[H]
\centering
\includegraphics[width=0.450\textwidth,height=0.220\textheight]{System_Model.eps}\\
\caption{System model for a hybrid IRS-aided AF relay wireless network.}\label{System_Model}
\end{figure}
The sets of $N$ units, $K$ active units and $L$ passive units can be respectively denoted as $\cal E_N$, $\cal E_K$ and $\cal E_L$, while $\cal E_N=\cal E_K\cup\cal E_L$ and $\cal E_K\cap\cal E_L=\varnothing$. $\boldsymbol\Theta$, $\boldsymbol\Phi$ and $\boldsymbol\Psi$ are the reflecting coefficient matrices corresponding to $\cal E_N$, $\cal E_K$ and $\cal E_L$. Further, we have $\boldsymbol\Theta=\boldsymbol\Phi+\boldsymbol\Psi$, where $\boldsymbol\Theta = \text{diag}( \alpha_1, \cdots,\alpha_N )$, $\boldsymbol\Phi = \text{diag}( \phi_1, \cdots, \phi_N )$ and $\boldsymbol\Psi = \text{diag}( \psi_1, \cdots, \psi_N )$. The reflecting coefficients of $i$th element in $\boldsymbol\Theta$, $\boldsymbol\Phi$ and $\boldsymbol\Psi$ are respectively expressed by
\begin{subnumcases}{\alpha_i=}{}
|\beta_i| e^{j\theta _{i}},~~~~~~~~~i\in \cal E_K, \\
e^{j\theta _{i}},~~~~~~~~~~~~~i\in \cal E_L,
\end{subnumcases}
\begin{subnumcases}{\phi_i=}{}
|\beta_i| e^{j\theta _{i}},~~~~~~~~~i\in \cal E_K, \\
0,~~~~~~~~~~~~~~~~~i\in \cal E_L,
\end{subnumcases}
\begin{subnumcases}{\psi_i=}{}
0,~~~~~~~~~~~~~~~~i\in \cal E_K, \\
e^{j\theta _{i}},~~~~~~~~~~~~~i\in \cal E_L,
\end{subnumcases}
where $|\beta_i|$ and $\theta _{i}\in ( 0 ,{2\pi } ]$ are the amplifying coefficient and the phase shift of the $i$th unit.
For convenience, let us define
\begin{equation}
\boldsymbol\Phi=\textbf{E}_K\boldsymbol\Theta, ~~~\boldsymbol\Psi=\overline{\textbf{E}}_K\boldsymbol\Theta,
\end{equation}
where
$\textbf{E}_K\in \mathbb R^{N \times N}$ and $\overline{\textbf{E}}_K\in \mathbb R^{N \times N}$ are diagonal matrices, which satisfy $\textbf{E}_K+\overline{\textbf{E}}_K=\textbf{I}_N$ and $\textbf{E}_K\overline{\textbf{E}}_K=\textbf{0}_N$. For $\textbf{E}_K$, 0 and 1 on the diagonals are respectively corresponding to the passive and active units, while $\overline{\textbf{E}}_K$ is opposite to $\textbf{E}_K$. It is assumed that the direct channel between S and D is blocked, and the signal reflected twice or more via IRS can be regardless of due to weak power.
All links are assumed to follow Rayleigh fading, the links from S to IRS and AF relay, from IRS to AF relay, from AF relay to IRS and D, from IRS to D are represented as
$\textbf{h}_{si}\in \mathbb C^{N \times 1}$,
$\textbf{h}_{sr}\in \mathbb C^{M \times 1}$,
$\textbf{H}_{ir}\in \mathbb C^{M \times N}$
$\textbf{H}_{ir}^H\in \mathbb C^{N \times M}$,
$\textbf{h}_{rd}^H\in \mathbb C^{1 \times M}$
and $\textbf{h}_{id}^H\in \mathbb C^{1 \times N}$, respectively.

In the first time slot, the received signal at IRS is denoted as
\begin{equation}\label{y_1i_r}
\textbf{y}_{1i}^r=\sqrt {{P_s}}\textbf{h}_{si}x+\textbf{n}_{1i},
\end{equation}
where $x$ with $\mathbb{E}\{x^H{x}\}=1$ and $P_s$ are the signal and power transmitted by S, respectively. $\textbf{n}_{1i}\sim \mathcal{CN}( \textbf{0},\sigma_{1i}^2\textbf{E}_K{\bf I}_{N} )$ caused by $K$ active elements represents the additive white Gaussian noise (AWGN). The received signal at AF relay is denoted as
\begin{align}\label{y_r}
&\textbf{y}_r=\sqrt {P_s}\textbf{h}_{sr}x + \sqrt {P_s}\textbf{H}_{ir}\boldsymbol\Theta_1\textbf{h}_{si}x + \textbf{H}_{ir}\boldsymbol\Phi_1\textbf{n}_{1i} + \textbf{n}_r \nonumber\\
&~~~ =\sqrt {P_s}(\textbf{h}_{sr} + \textbf{H}_{ir}\boldsymbol\Theta_1\textbf{h}_{si})x + \textbf{H}_{ir}\textbf{E}_K\boldsymbol\Theta_1\textbf{n}_{1i} + \textbf{n}_r,
\end{align}
where $\boldsymbol\Theta_1 = \text{diag}( \alpha_{11}, \cdots,\alpha_{1N} )$ and $\boldsymbol\Phi_1 = \text{diag}( \phi_{11}, \cdots, \phi_{1N} )$ are the reflecting coefficient matrices of $\cal E_N$ and $\cal E_K$ in the first time slot. $\textbf{n}_r$ is the AWGN subject to the distribution of $\textbf{n}_r\sim \mathcal{CN}( \textbf{0},\sigma_r^2{\bf I}_{M} )$.

In the second time slot, the transmit signal from AF relay is given by
\begin{equation}\label{y_t}
\textbf{y}_t=\textbf{A}\textbf{y}_r,
\end{equation}
where $\textbf{A}\in \mathbb C^{M \times M}$ is the beamforming matrix . The reflected signal from IRS is expressed as
\begin{equation}\label{y_2i}
\textbf{y}_{2i}^t=\boldsymbol\Theta_2\textbf{H}_{ir}^H\textbf{y}_t+\boldsymbol\Phi_2\textbf{n}_{2i},
\end{equation}
where $\boldsymbol\Theta_2 = \text{diag}( \alpha_{21}, \cdots,\alpha_{2N} )$ and $\boldsymbol\Phi_2 = \text{diag}( \phi_{21}, \cdots, \phi_{2N} )$ are the reflecting coefficient matrices of $\cal E_N$ and $\cal E_K$ in the second time slot. $\textbf{n}_{2i}$ is the AWGN, which obeys the distribution of $\textbf{n}_{2i}\sim \mathcal{CN}( \textbf{0},\sigma_{2i}^2\textbf{E}_K{\bf I}_{N} )$. The received signal at D is given by
\begin{align}\label{y_d}
y_d&=(\textbf{h}_{rd}^H + \textbf{h}_{id}^H\boldsymbol\Theta_2\textbf{H}_{ir}^H)\textbf{y}_t + \textbf{h}_{id}^H\boldsymbol\Phi_2\textbf{n}_{2i} + \text{n}_d \nonumber\\
%&=(\textbf{h}_{rd}^H + \textbf{h}_{id}^H\boldsymbol\Theta_2\textbf{H}_{ir}^H)\textbf{y}_t + \textbf{h}_{id}^H\textbf{E}_K\boldsymbol\Theta_2\textbf{n}_{2i} + \text{n}_d, \\
&=\sqrt {P_s}(\textbf{h}_{rd}^H + \textbf{h}_{id}^H\boldsymbol\Theta_2\textbf{H}_{ir}^H)\textbf{A}(\textbf{h}_{sr} +
\textbf{H}_{ir}\boldsymbol\Theta_1\textbf{h}_{si})x \nonumber\\
&~~~~~~~+ (\textbf{h}_{rd}^H + \textbf{h}_{id}^H\boldsymbol\Theta_2\textbf{H}_{ir}^H)\textbf{A}( \textbf{H}_{ir}\textbf{E}_K\boldsymbol\Theta_1\textbf{n}_{1i} + \textbf{n}_r ) + \textbf{h}_{id}^H\textbf{E}_K\boldsymbol\Theta_2\textbf{n}_{2i} + \text{n}_d,
\end{align}
where $\text{n}_d$ is the AWGN following the distribution of $\text{n}_d\sim \mathcal{CN}( 0,\sigma_d^2 )$.
%Substituting (\ref{y_r}) and (\ref{y_t}) into (\ref{y_d}) yields
%\begin{align}\label{y_d1}
%&y_d=\sqrt {P_s}(\textbf{h}_{rd}^H + \textbf{h}_{id}^H\boldsymbol\Theta_2\textbf{H}_{ir}^H)\textbf{A}(\textbf{h}_{sr} +
%\textbf{H}_{ir}\boldsymbol\Theta_1\textbf{h}_{si})x \nonumber\\
%&~~~~~~~+ (\textbf{h}_{rd}^H + \textbf{h}_{id}^H\boldsymbol\Theta_2\textbf{H}_{ir}^H)\textbf{A}( \textbf{H}_{ir}\textbf{E}_K\boldsymbol\Theta_1\textbf{n}_{1i} + \textbf{n}_r ) + \textbf{h}_{id}^H\textbf{E}_K\boldsymbol\Theta_2\textbf{n}_{2i} + \text{n}_d.
%\end{align}
Let us define $ \sigma_{1i}^2=\sigma_{2i}^2=\sigma_{r}^2=\sigma_d^2=\sigma^2 $ and $\gamma_s=\frac{P_s}{\sigma^2} $, we have achievable system rate
\begin{equation}\label{R}
R=\frac{1}{2}\log_2 (1 + \text{SNR}),
\end{equation}
where SNR is represented as
\begin{equation}\label{SNR_1}
\text{SNR}=\frac{\gamma_s|(\textbf{h}_{rd}^H + \textbf{h}_{id}^H\boldsymbol\Theta_2\textbf{H}_{ir}^H)\textbf{A}(\textbf{h}_{sr} +
          \textbf{H}_{ir}\boldsymbol\Theta_1\textbf{h}_{si})|^2}  {\|(\textbf{h}_{rd}^H + \textbf{h}_{id}^H\boldsymbol\Theta_2\textbf{H}_{ir}^H)\textbf{A}\textbf{H}_{ir}\textbf{E}_K\boldsymbol\Theta_1\|^2+ \|(\textbf{h}_{rd}^H + \textbf{h}_{id}^H\boldsymbol\Theta_2\textbf{H}_{ir}^H)\textbf{A}\|^2+ \|\textbf{h}_{id}^H\textbf{E}_K\boldsymbol\Theta_2\|^2 + 1}.
\end{equation}

\subsection{Problem Formulation}
Similar to \cite{2022WXH}, it is assumed that all the channel state informations for the proposed hybrid IRS-aided AF relay wireless network can be available. Then the optimization problem is designed as
\begin{subequations}\label{OP}
\begin{align}
&\max \limits_{\boldsymbol\Theta_1, \boldsymbol\Theta_2, \textbf{A} }~~~ \text{SNR} \\
&~~~\text{s.t.}~~ ~~~  |\boldsymbol\Theta_1(i,i)|=1,~|\boldsymbol\Theta_2(i,i)|=1,~\text{for}~i\in \cal E_L,  \label{OP_4}\\
%&~~~~~~~~  |\boldsymbol\Theta_1(i,i)|>1,~|\boldsymbol\Theta_2(i,i)|>1,~\text{for}~i\in \cal E_K,  \label{OP_5}\\
&~~~~~~~~~~~  \gamma_s\|\textbf{E}_K\boldsymbol\Theta_1\textbf{h}_{si}\|^2+\|\textbf{E}_K\boldsymbol\Theta_1\|_F^2\leq \gamma_i,\label{OP_1}\\
&~~~~~~~~~~~  \gamma_s\|\textbf{A}(\textbf{h}_{sr} + \textbf{H}_{ir}\boldsymbol\Theta_1\textbf{h}_{si})\|^2   + \|\textbf{A}\textbf{H}_{ir}\textbf{E}_K\boldsymbol\Theta_1\|_F^2+ \|\textbf{A}\|_F^2\leq \gamma_r,\label{OP_2}\\
&~~~~~~~~~~~  \gamma_s\|\textbf{E}_K\boldsymbol\Theta_2\textbf{H}_{ir}^H\textbf{A}(\textbf{h}_{sr} +\textbf{H}_{ir}\boldsymbol\Theta_1\textbf{h}_{si})\|^2  + \|\textbf{E}_K\boldsymbol\Theta_2\textbf{H}_{ir}^H\textbf{A}\textbf{H}_{ir}\textbf{E}_K\boldsymbol\Theta_1\|_F^2 \nonumber\\
&~~~~~~~~~~~  + \|\textbf{E}_K\boldsymbol\Theta_2\textbf{H}_{ir}^H\textbf{A}\|_F^2+\|\textbf{E}_K\boldsymbol\Theta_2\|_F^2\leq \gamma_i,  \label{OP_3}
%&~~~~~~~~  \beta_{1i}>1,~~~ \beta_{2i}>1,
\end{align}
\end{subequations}
where $\gamma_i=\frac{P_i}{\sigma^2} $ and $\gamma_r=\frac{P_r}{\sigma^2} $, $P_i$ and $P_r$ are the transmit power budgets of hybrid IRS and AF relay. Since the IRS is hybrid consisting of active and passive elements, it is difficult to solve the optimization problem. To enhance the rate performance, two efficient beamforming methods called HP-SDR-FP and WF-GPI-GRR are proposed to optimize AF relay beamforming matrix $\textbf{A}$, IRS reflecting coefficient matrices $\boldsymbol\Theta_1$ and $\boldsymbol\Theta_2$.

\section{Proposed a High-performance SDR-FP-based Max-SNR Method}\label{SDR_FP}
In this section, a HP-SDR-FP method is proposed to solve problem (\ref{OP}) for maximum SNR. To facilitate processing, problem (\ref{OP}) is decoupled into three subproblems by optimizing one and fixing the other two. For each subproblem, we firstly relax it as an SDR problem, and combine Charnes-Cooper transformation of FP algorithm to solve the SDR problem. Furthermore, Gaussian randomization method is applied to recover the rank-1 solution.
\subsection{Optimization of $\textbf{A}$ Given $\bf\Theta_1$ and $\bf\Theta_2$}
Given $\boldsymbol\Theta_1$ and $\boldsymbol\Theta_2$, the optimization problem is reduced to
\begin{subequations}
\begin{align}
&\max \limits_{ \textbf{A} } ~~~~\text{SNR}   \\
&~\text{s.t.}~~~~~~\text{(\ref{OP_2})},~~~\text{(\ref{OP_3})}.
\end{align}
\end{subequations}
Let us define $\textbf{a}=\text{vec}(\textbf{A})\in \mathbb C^{M^2 \times 1}$,  SNR can be translated to
\begin{equation}\label{SNR_2}
\text{SNR}=\frac{ \gamma_s\textbf{a}^H\textbf{B}_1\textbf{a} } {\textbf{a}^H(\textbf{B}_2+\textbf{B}_3)\textbf{a}+  \|\textbf{h}_{id}^H\textbf{E}_K\boldsymbol\Theta_2\|^2 + 1 },
\end{equation}
where
$\textbf{B}_1=[(\textbf{h}_{sr} + \textbf{H}_{ir}\boldsymbol\Theta_1\textbf{h}_{si})^*(\textbf{h}_{sr} + \textbf{H}_{ir}\boldsymbol\Theta_1\textbf{h}_{si})^T]\otimes[(\textbf{h}_{rd}^H + \textbf{h}_{id}^H\boldsymbol\Theta_2\textbf{H}_{ir}^H)^H(\textbf{h}_{rd}^H + \textbf{h}_{id}^H\boldsymbol\Theta_2\textbf{H}_{ir}^H)]$,
$\textbf{B}_2=[(\textbf{H}_{ir}\textbf{E}_K\boldsymbol\Theta_1)^*(\textbf{H}_{ir}\textbf{E}_K\boldsymbol\Theta_1)^T]\\ \otimes[(\textbf{h}_{rd}^H + \textbf{h}_{id}^H\boldsymbol\Theta_2\textbf{H}_{ir}^H)^H(\textbf{h}_{rd}^H + \textbf{h}_{id}^H\boldsymbol\Theta_2\textbf{H}_{ir}^H)]$ and
$\textbf{B}_3=\textbf{I}_{M}\otimes[(\textbf{h}_{rd}^H + \textbf{h}_{id}^H\boldsymbol\Theta_2\textbf{H}_{ir}^H)^H( \textbf{h}_{rd}^H + \textbf{h}_{id}^H\boldsymbol\Theta_2\textbf{H}_{ir}^H)]$.
%SNR can be rewritten as
%\begin{equation}
%\text{SNR}=\frac{ \gamma_s\text{tr}( \textbf{B}_1\widehat{\textbf{A}} ) } {\text{tr}\{( \textbf{B}_2+\textbf{B}_3 )\widehat{\textbf{A}}\}+  \left\|\textbf{h}_{id}^H\textbf{E}_K\boldsymbol\Theta_2\right\|^2 + 1 }.
%\end{equation}
In the same manner, the constraints (\ref{OP_2}) and (\ref{OP_3}) can be respectively converted to
\begin{subequations}
\begin{align}
&\textbf{a}^H(\gamma_s\textbf{C}_1+\textbf{C}_2+\textbf{I}_{M^2})\textbf{a}\leq\gamma_r,\label{OP_2_1} \\
&\textbf{a}^H(\gamma_s\textbf{D}_1+\textbf{D}_2+\textbf{D}_3)\textbf{a}+\|\textbf{E}_K\boldsymbol\Theta_2\|_F^2\leq \gamma_i,\label{OP_3_1}
\end{align}
\end{subequations}
where
$\textbf{C}_1=[(\textbf{h}_{sr} + \textbf{H}_{ir}\boldsymbol\Theta_1\textbf{h}_{si})^*(\textbf{h}_{sr} + \textbf{H}_{ir}\boldsymbol\Theta_1\textbf{h}_{si})^T]\otimes\textbf{I}_M$,
$\textbf{C}_2=[(\textbf{H}_{ir}\textbf{E}_K\boldsymbol\Theta_1)^*(\textbf{H}_{ir}\textbf{E}_K\boldsymbol\Theta_1)^T]\otimes\textbf{I}_M$,
$\textbf{D}_1=[(\textbf{h}_{sr} + \textbf{H}_{ir}\boldsymbol\Theta_1\textbf{h}_{si})^*(\textbf{h}_{sr} + \textbf{H}_{ir}\boldsymbol\Theta_1\textbf{h}_{si})^T] \otimes[(\textbf{E}_K\boldsymbol\Theta_2\textbf{H}_{ir}^H)^H(\textbf{E}_K\boldsymbol\Theta_2\textbf{H}_{ir}^H)]$,
$\textbf{D}_2=[(\textbf{H}_{ir}\textbf{E}_K\boldsymbol\Theta_1)^*(\textbf{H}_{ir}\textbf{E}_K\boldsymbol\Theta_1)^T] \otimes[(\textbf{E}_K\boldsymbol\Theta_2\textbf{H}_{ir}^H)^H(\textbf{E}_K\boldsymbol\Theta_2\textbf{H}_{ir}^H )]$ and
$\textbf{D}_3=\textbf{I}_M\otimes[(\textbf{E}_K\boldsymbol\Theta_2\textbf{H}_{ir}^H)^H(\textbf{E}_K\boldsymbol\Theta_2\textbf{H}_{ir}^H)]$.
Let us define $\widehat{\textbf{A}}=\textbf{a}\textbf{a}^H\in \mathbb C^{M^2 \times M^2}$, in accordance with the rank inequality: $\text{rank}~\textbf{P} \leq \text{min}\{m, n\}$, where $\textbf{P}\in \mathbb C^{m \times n}$, we can get $\text{rank}(\widehat{\textbf{A}})\leq\text{rank}(\textbf{a})=1$. The optimization problem can be recast as
\begin{subequations}\label{rank_A}
\begin{align}
&\max \limits_{ \widehat{\textbf{A}} } ~~~\frac{ \gamma_s\text{tr}( \textbf{B}_1\widehat{\textbf{A}} ) } {\text{tr}\{( \textbf{B}_2+ \textbf{B}_3 )\widehat{\textbf{A}}\}+  \|\textbf{h}_{id}^H\textbf{E}_K\boldsymbol\Theta_2\|^2 + 1 } \label{rank_A_1}\\
&~\text{s.t.}~~~~~\text{tr}\{(\gamma_s\textbf{C}_1+\textbf{C}_2+\textbf{I}_{M^2})\widehat{\textbf{A}}\}\leq\gamma_r, \label{rank_A_2}\\
&~~~~~~~~~~ \text{tr}\{(\gamma_s\textbf{D}_1+\textbf{D}_2+\textbf{D}_3)\widehat{\textbf{A}}\}+\|\textbf{E}_K\boldsymbol\Theta_2\|_F^2\leq \gamma_i,  \label{rank_A_3}\\
&~~~~~~~~~~ \widehat{\textbf{A}}\succeq\textbf{0},~~~\text{rank}(\widehat{\textbf{A}})=1,
\end{align}
\end{subequations}
which is a non-convex problem because of rank-one constraint. After removing $\text{rank}(\widehat{\textbf{A}})=1$ constraint, we have the SDR problem of (\ref{rank_A}) as follows
\begin{subequations}\label{A}
\begin{align}
&\max \limits_{ \widehat{\textbf{A}} } ~~~\frac{ \gamma_s\text{tr}( \textbf{B}_1\widehat{\textbf{A}} ) } {\text{tr}\{( \textbf{B}_2+\textbf{B}_3 )\widehat{\textbf{A}}\}+  \|\textbf{h}_{id}^H\textbf{E}_K\boldsymbol\Theta_2\|^2 + 1 } \label{A_1}\\
&~\text{s.t.}~~~~~\text{(\ref{rank_A_2})},~~~\text{(\ref{rank_A_3})},~~~\widehat{\textbf{A}}\succeq\textbf{0}.
\end{align}
\end{subequations}
The objective function (\ref{A_1}) is a linear fractional function with respect to $\widehat{\textbf{A}}$, which is a quasi-convex function with the denominator $>0$, so problem (\ref{A}) is a quasi-convex problem with convex constraints. It is necessary to apply Charnes-Cooper transformation, which helps convert the optimization problem from quasi-convex to convex. Introducing a slack variable $m$ and defining $m=(\text{tr}\{( \textbf{B}_2+\textbf{B}_3 )\widehat{\textbf{A}}\}+  \|\textbf{h}_{id}^H\textbf{E}_K\boldsymbol\Theta_2\|^2 + 1)^{-1}$, the above problem (\ref{A}) is further rewritten as follows
\begin{subequations}\label{t}
\begin{align}
&\max \limits_{ \widetilde{\textbf{A}},m }~~~  \gamma_s\text{tr}\{\textbf{B}_1\widetilde{\textbf{A}}\} \\
&~\text{s.t.}~~~~  \text{tr}\{(\gamma_s\textbf{C}_1+\textbf{C}_2+\textbf{I}_{M^2})\widetilde{\textbf{A}}\}\leq m\gamma_r,\\
&~~~~~~~~~ \text{tr}\{(\gamma_s\textbf{D}_1+\textbf{D}_2+\textbf{D}_3)\widetilde{\textbf{A}}\}+m\|\textbf{E}_K\boldsymbol\Theta_2\|_F^2\leq m\gamma_i, \\
&~~~~~~~~~ \text{tr}\{( \textbf{B}_2+\textbf{B}_3 )\widetilde{\textbf{A}}\}+m\|\textbf{h}_{id}^H\textbf{E}_K\boldsymbol\Theta_2\|^2 + m=1,\\
&~~~~~~~~~ \widetilde{\textbf{A}}\succeq\textbf{0},~m>0,
\end{align}
\end{subequations}
where $\widetilde{\textbf{A}}=m\widehat{\textbf{A}}$. Clearly, the above optimization problem has become a SDP problem, which is directly solved by CVX. The solution to problem (\ref{A}) is $\widehat{\textbf{A}}=\widetilde{\textbf{A}}/m$. However, the rank-one constraint $\text{rank}(\widehat{\textbf{A}})=1$ is not considered in the SDR problem. Since the obtained solution $\widehat{\textbf{A}}$ is not generally rank-one matrix, the Gaussian randomization method is applied to achieve a rank-one solution $\widehat{\textbf{A}}$,
%and a rank-one solution $\widehat{\textbf{A}}$ is obtained, Additionally, $\textbf{a}$ is obtained by performing eigenvalue decomposition on $\widehat{\textbf{A}}$,
thereby, AF relay beamforming matrix $\textbf{A}$ is achieved.

\subsection{Optimization of $\bf\Theta_1$ Given $\textbf{A}$ and $\bf\Theta_2$}
Given that $\textbf{A}$ and $\boldsymbol\Theta_2$ are fixed, the optimization problem can be represented by as follows
\begin{subequations}\label{givenA_Theta_1}
\begin{align}
&\max  \limits_{{\boldsymbol\Theta_1} } ~~ \frac{\gamma_s|\textbf{h}_{rid}^H(\textbf{h}_{sr} +\textbf{H}_{ir}\boldsymbol\Theta_1\textbf{h}_{si})|^2} {\|\textbf{h}_{rid}^H\textbf{H}_{ir}\textbf{E}_K\boldsymbol\Theta_1\|^2+ \|\textbf{h}_{rid}^H\|^2+ \|\textbf{h}_{id}^H\textbf{E}_K\boldsymbol\Theta_2\|^2 + 1} \label{givenA_Theta_1_1}\\
&~\text{s.t.}~~~~   |\boldsymbol\Theta_1(i,i)|=1,~~\text{for}~i\in \cal E_L,  \label{givenA_Theta_1_2}\\
%&~~~~~~~~|\boldsymbol\Theta_1(i,i)|>1,~~\text{for}~i\in \cal E_K,  \label{givenA_Theta_1_3}\\
%&~~~~~~~~~~~|(\textbf{E}_K\boldsymbol\Theta_1)(i,i)|=0,~~\text{for}~i\in \cal E_L,  \label{givenA_Theta_1_3}\\
%&~~~~~~~~~~~|(\textbf{E}_K\boldsymbol\Theta_1)(i,i)|=\beta_{1i},~~\text{for}~i\in \cal E_K,  \label{givenA_Theta_1_4}\\
%&~~~~~~~~\gamma_s\left\|\textbf{A}(\textbf{h}_{sr} + \textbf{H}_{ir}\boldsymbol\Theta_1\textbf{h}_{si})\right\|^2\nonumber\\
%&~~~~~~~~~~~~~~~~~~~~~~+ \left\|\textbf{A}\textbf{H}_{ir}\textbf{E}_K\boldsymbol\Theta_1\right\|_F^2+ \left\|\textbf{A}\right\|_F^2\leq \gamma_r, \label{givenA_Theta_1_5} \\
&~~~~~~~~~\text{(\ref{OP_1})},~~~ \text{(\ref{OP_2})}, ~~~\text{(\ref{OP_3})},
\end{align}
\end{subequations}
where $\textbf{h}_{rid}=[(\textbf{h}_{rd}^H + \textbf{h}_{id}^H\boldsymbol\Theta_2\textbf{H}_{ir}^H)\textbf{A}]^H$.
In order to further simplify the objective function and constraints of the optimization problem, let us define $\textbf{u}_1=[ \alpha_{11}, \cdots, \alpha_{1N}]^T$, we have $\textbf{h}_{sr} + \textbf{H}_{ir}\boldsymbol\Theta_1\textbf{h}_{si}=\textbf{H}_{sir}\textbf{v}_1$ and $\textbf{h}_{rid}^H\textbf{H}_{ir}\textbf{E}_K\boldsymbol\Theta_1=\textbf{u}_1^T\text{diag}\{\textbf{h}_{rid}^H\textbf{H}_{ir}\textbf{E}_K\}$, where $\textbf{v}_1=[ \textbf{u}_1; 1]$ and $\textbf{H}_{sir}=[\textbf{H}_{ir}\text{diag}\{\textbf{h}_{si}\}, \textbf{h}_{sr}]$.
Substituting these formulas into (\ref{givenA_Theta_1_1}), and due to the fact that $\|\textbf{u}_1^T\text{diag}\{\textbf{h}_{rid}^H\textbf{H}_{ir}\textbf{E}_K \}\|^2\\=\|\text{diag}\{\textbf{h}_{rid}^H\textbf{H}_{ir}\textbf{E}_K\}\textbf{u}_1\|^2$, the object function can be further rewritten as
\begin{equation}\label{Oj1}
\frac{\textbf{v}_1^H\textbf{F}_1\textbf{v}_1}{\textbf{v}_1^H\textbf{F}_2\textbf{v}_1 },
\end{equation}
where
$\textbf{F}_1=\gamma_s\textbf{H}_{sir}^H\textbf{h}_{rid}\textbf{h}_{rid}^H\textbf{H}_{sir}$ and
$\textbf{F}_2=[
\text{diag}\{\textbf{E}_K\textbf{H}_{ir}^H\textbf{h}_{rid}\}\text{diag}\{\textbf{h}_{rid}^H\textbf{H}_{ir}\textbf{E}_K\},~ \textbf{0}_{N \times 1};~ \textbf{0}_{1 \times N},~ \|\textbf{h}_{rid}^H\|^2+ \|\textbf{h}_{id}^H\textbf{E}_K\boldsymbol\Theta_2\|^2 + 1]$.
The constraint (\ref{givenA_Theta_1_2}) for passive elements $\cal E_L$ can be rewritten as
\begin{equation}
%\begin{align}
|\textbf{v}_1(i)|^2=1,~~\text{for}~i\in \cal E_L. \label{SP1_1}
%&|\textbf{v}_1(i)|^2>1,~~\text{for}~i\in \cal E_K. \label{SP1_2}
%\end{align}
\end{equation}
Obviously, $\|\textbf{E}_K\boldsymbol\Theta_1\|_F^2=\|\textbf{E}_K\textbf{u}_1\|^2$, and the constraint (\ref{OP_1}) can be translated to
\begin{equation}\label{vHv_1}
\textbf{v}_1^H\textbf{G}_1\textbf{v}_1\leq\gamma_i,
\end{equation}
where
$\textbf{G}_1=[
\gamma_s\text{diag}\{\textbf{h}_{si}^H\}\textbf{E}_K\text{diag}\{\textbf{h}_{si}\}+ \textbf{E}_K,~ \textbf{0}_{N \times 1};~
\textbf{0}_{1 \times N},~ \text{0}]$.
%Then for constraint (\ref{OP_2}), a variable $\tau_1$ is introduced to Cauchy-Schwartz inequality, which yields
%\begin{equation}\label{Cauchy_Schwartz}
%\left\|\textbf{A}\textbf{H}_{ir}\textbf{E}_K\boldsymbol\Theta_1\right\|_F^2=\tau_1\left\|\textbf{A}\textbf{H}_{ir}\right\|_F^2\left\|\textbf{E}_K\boldsymbol\Theta_1\right\|_F^2.
%\end{equation}
Then for constraint (\ref{OP_2}), according to the property of Hadamard product: $ \text{tr}(\textbf{X}(\textbf{Y}\odot\textbf{Z}))= \text{tr}((\textbf{X}\odot\textbf{Y}^T)\textbf{Z})$, where $\textbf{X}\in \mathbb C^{m \times n}$, $\textbf{Y}\in \mathbb C^{n \times m}$ and $\textbf{Z}\in \mathbb C^{n \times m}$, we have
\begin{align}\label{Cauchy_Schwartz}
\|\textbf{A}\textbf{H}_{ir}\textbf{E}_K\boldsymbol\Theta_1\|_F^2&=\|\textbf{A}\textbf{H}_{ir}\textbf{E}_K\text{diag}\{ \textbf{u}_1\}\|_F^2\nonumber\\
&=\text{tr}\{ \textbf{E}_K\textbf{H}_{ir}^H\textbf{A}^H\textbf{A}\textbf{H}_{ir}\textbf{E}_K\text{diag}\{\textbf{u}_1\}\text{diag}\{ \textbf{u}_1^H\} \}\nonumber\\
&=\text{tr}\{ \textbf{E}_K\textbf{H}_{ir}^H\textbf{A}^H\textbf{A}\textbf{H}_{ir}[\textbf{E}_K\odot(\textbf{u}_1\textbf{u}^H_1)]\}\nonumber\\
&= \textbf{u}^H_1[(\textbf{E}_K\textbf{H}_{ir}^H\textbf{A}^H\textbf{A}\textbf{H}_{ir})\odot\textbf{E}_K]\textbf{u}_1\nonumber\\
&= \textbf{u}^H_1[(\textbf{H}_{ir}^H\textbf{A}^H\textbf{A}\textbf{H}_{ir})\odot\textbf{E}_K]\textbf{u}_1.
\end{align}
Inserting $\textbf{h}_{sr} + \textbf{H}_{ir}\boldsymbol\Theta_1\textbf{h}_{si}=\textbf{H}_{sir}\textbf{v}_1$ and (\ref{Cauchy_Schwartz}) back into the constraint (\ref{OP_2}), which can be rewritten as
%\begin{align}\label{gamma_r}
%&\gamma_s\left\|\textbf{A}(\textbf{h}_{sr} + \textbf{H}_{ir}\boldsymbol\Theta_1\textbf{h}_{si})\right\|^2+\left\|\textbf{A}\textbf{H}_{ir}\textbf{E}_K\boldsymbol\Theta_1\right\|^2+\left\|\textbf{A}\right\|^2 \nonumber\\
%&=\textbf{v}_1^H\left\{\gamma_s\textbf{H}_{sir}^H\textbf{A}^H\textbf{A}\textbf{H}_{sir}+\left[
%\begin{array}{cccc}
%\tau\left\|\textbf{A}\textbf{H}_{ir}\textbf{E}_K\right\|^2\textbf{I}_N & \textbf{0}_{N \times 1}\\
%\textbf{0}_{1 \times N} & \left\|\textbf{A}\right\|^2
%\end{array}
%\right]\right\}\textbf{v}_1 \nonumber\\
%&=\textbf{v}_1^H\textbf{H}_{sr}\textbf{v}_1 \nonumber\\
%&\leq\gamma_r
%\end{align}
\begin{equation}\label{vHv_2}
\textbf{v}_1^H\textbf{G}_2\textbf{v}_1\leq\gamma_r,
\end{equation}
where
$\textbf{G}_2=\gamma_s\textbf{H}_{sir}^H\textbf{A}^H\textbf{A}\textbf{H}_{sir} +[
(\textbf{H}_{ir}^H\textbf{A}^H\textbf{A}\textbf{H}_{ir})\odot\textbf{E}_K,~ \textbf{0}_{N \times 1};~
\textbf{0}_{1 \times N},~ \|\textbf{A}\|_F^2]$.
Similarly, (\ref{OP_3}) can be written in the following form
\begin{equation}\label{vHv_3}
\textbf{v}_1^H\textbf{G}_3\textbf{v}_1\leq\gamma_i,
\end{equation}
where
$\textbf{G}_3=\gamma_s\textbf{H}_{sir}^H\textbf{A}^H\textbf{H}_{ir}\boldsymbol\Theta_2^H\textbf{E}_K\boldsymbol\Theta_2\textbf{H}_{ir}^H\textbf{A}\textbf{H}_{sir}
+[
(\textbf{H}_{ir}^H\textbf{A}^H\textbf{H}_{ir}\boldsymbol\Theta_2^H\textbf{E}_K\boldsymbol\Theta_2\textbf{H}_{ir}^H\textbf{A}\textbf{H}_{ir})\odot\textbf{E}_K ,~ \textbf{0}_{N \times 1};~
\textbf{0}_{1 \times N},~\\ \|\textbf{E}_K\boldsymbol\Theta_2\textbf{H}_{ir}^H\textbf{A}\|_F^2+\|\textbf{E}_K\boldsymbol\Theta_2\|_F^2
]$.
Substituting the simplified objective function and constraints into (\ref{givenA_Theta_1}), the optimization problem can be equivalently transformed into
\begin{subequations}\label{v1Hv1}
\begin{align}
&~\max \limits_{\textbf{v}_1}~~~ \text{(\ref{Oj1})} \\
&~~\text{s.t.}~~~~~  \text{(\ref{SP1_1})},~\text{(\ref{vHv_1})},~\text{(\ref{vHv_2})},~\text{(\ref{vHv_3})},~\textbf{v}_1(N+1)=1.
\end{align}
\end{subequations}
Aiming at further transforming the optimization problem, and defining $\textbf{V}_1=\textbf{v}_1\textbf{v}_1^H$, problem (\ref{v1Hv1}) can be equivalently given by
\begin{subequations}\label{trV1}
\begin{align}
&\max \limits_{\textbf{V}_1 } ~~~~\frac{\text{tr}(\textbf{F}_1\textbf{V}_1)}{\text{tr}(\textbf{F}_2\textbf{V}_1) } \\
&~\text{s.t.} ~~~~~~\textbf{V}_1(i,i)=1, ~~\text{for}~i\in \cal E_L,\\
%&~~~~~~~~~~  \textbf{V}_1(i,i)>1, ~~\text{for}~i\in \cal E_K,\\
&~~~~~~~~~~~  \textbf{V}_1(N+1,N+1)=1, \\
&~~~~~~~~~~~  \text{tr}(\textbf{G}_1\textbf{V}_1)\leq\gamma_i,~\text{tr}(\textbf{G}_2\textbf{V}_1)\leq\gamma_r, \\
&~~~~~~~~~~~  \text{tr}(\textbf{G}_3\textbf{V}_1)\leq\gamma_i,~\text{rank}(\textbf{V}_1)=1,~\textbf{V}_1\succeq\textbf{0}.
\end{align}
\end{subequations}
Due to the fact that the object function is quasi-convex and constraint $\text{rank}(\textbf{V}_1)=1$ is non-convex, (\ref{trV1}) is still a non-convex problem.
Relaxing the rank-1 constraint, the problem is transformed into a SDR problem, which can also be solved by applying Charnes-Cooper transformation. Moreover, introducing a slack variable $\tau$, then defining $\tau=\text{tr}(\textbf{F}_2\textbf{V}_1)^{-1}$ and $\widetilde{\textbf{V}}_1=\tau\textbf{V}_1$, the SDR problem of (\ref{trV1}) can be translated to a SDP problem, i.e.
\begin{subequations}\label{SDP}
\begin{align}
&\max \limits_{\widetilde{\textbf{V}}_1, \tau } ~~~~~ \text{tr}(\textbf{F}_1\widetilde{\textbf{V}}_1) \\
&~\text{s.t.}~~~~~~~ \widetilde{\textbf{V}}_1(i,i)=\tau, ~~\text{for}~i\in \cal E_L,\\
%&~~~~~~~~~~~ \widetilde{\textbf{V}}_1(i,i)>\tau, ~~\text{for}~i\in \cal E_K,\\
&~~~~~~~~~~~~ \widetilde{\textbf{V}}_1(N+1,N+1)=\tau,~\tau>0, \\
&~~~~~~~~~~~~ \text{tr}(\textbf{G}_1\widetilde{\textbf{V}}_1)\leq\tau\gamma_i,~\text{tr}(\textbf{G}_2\widetilde{\textbf{V}}_1)\leq\tau\gamma_r, \\
&~~~~~~~~~~~~ \text{tr}(\textbf{G}_3\widetilde{\textbf{V}}_1)\leq\tau\gamma_i,~\text{tr}(\textbf{F}_2\widetilde{\textbf{V}}_1)=1,~\widetilde{\textbf{V}}_1\succeq\textbf{0},
\end{align}
\end{subequations}
which can be directly solved by CVX, thereby the solution $\textbf{V}_1$ of SDR problem of (\ref{trV1}) is achieved, and Gaussian randomization method is used to recover a rank-one solution $\textbf{V}_1$.
%Then $\textbf{V}_1$ is performed the eigenvalue decomposition, i.e., $\textbf{V}_1=\textbf{Q}_1\mathbf{\Sigma_1}\textbf{Q}_1^H$, where $\textbf{Q}_1$ is the eigenvector matrix corresponding to the eigenvalue of $\textbf{V}_1$, $\mathbf{\Sigma_1}$ is a diagonal matrix consisted of eigenvalues. Since $\text{rank}(\textbf{V}_1)=1$, $\textbf{v}_1=\left(\sqrt{\mathbf{\Sigma_1}[1,1]}\textbf{Q}_1[1,:]\right)$.
Then the solution $\textbf{v}_1$ is extracted from the eigenvalue decomposition of $\textbf{V}_1$, subsequently, IRS reflecting coefficient matrix $\boldsymbol\Theta_1$ can be obtained as follows
%\begin{equation}
%\boldsymbol\Theta_1=\text{diag}\left(\left[\frac{\textbf{v}_1}{\textbf{v}_1(N+1)}\right](1:N)\right).
%\end{equation}
\begin{subnumcases}{\boldsymbol\Theta_1(i,i)=}{}
e^{j\text{arg}\left(\frac{\textbf{v}_1(i)}{\textbf{v}_1(N+1)}\right)},~~~~~~~~i\in \cal E_L, \\
\frac{\textbf{v}_1(i)}{\textbf{v}_1(N+1)},~~~~~~~~~~~~i\in \cal E_K.
\end{subnumcases}
\subsection{Optimization of $\bf\Theta_2$ Given $\textbf{A}$ and $\bf\Theta_1$}
In the subsection, defining
$\textbf{u}_2=[ \alpha_{21}, \cdots, \alpha_{2N}]^H$, we have
$\textbf{h}_{rd}^H + \textbf{h}_{id}^H\boldsymbol\Theta_2\textbf{H}_{ir}^H=\textbf{v}_2^H\textbf{H}_{rid}$ and
$\boldsymbol\Theta_2\textbf{H}_{ir}^H\textbf{A}(\textbf{h}_{sr} +\textbf{H}_{ir}\boldsymbol\Theta_1\textbf{h}_{si})=\text{diag}\{ \textbf{H}_{ir}^H\textbf{A}(\textbf{h}_{sr} +\textbf{H}_{ir}\boldsymbol\Theta_1\textbf{h}_{si})\} \textbf{u}_2^*$,
where $\textbf{v}_2=[ \textbf{u}_2; 1]$ and $\textbf{H}_{rid}=[\text{diag}\{\textbf{h}_{id}^H\}\textbf{H}_{ir}^H; \textbf{h}_{rd}^H]$.
%\begin{equation}
%\textbf{v}_2=[ \textbf{u}_2; 1],~\textbf{H}_{rid}=[\text{diag}\{\textbf{h}_{id}^H\}\textbf{H}_{ir}^H; \textbf{h}_{rd}^H].
%\end{equation}
When $\textbf{A}$ and $\boldsymbol\Theta_1$ are fixed, inserting the equivalent equations back into the optimization problem (\ref{OP}), which can be further transformed into as follows
%\begin{subequations}\label{givenA_Theta_2}
%\begin{align}
%&\max \limits_{\textbf{v}_2^H, \textbf{u}_2^H } \frac{\gamma_s\left|\textbf{v}_2^H\textbf{H}_{rid}\textbf{h}_{sir}\right|^2} {\left\|\textbf{v}_2^H\textbf{H}_{rid}\textbf{A}\textbf{H}_{ir}\boldsymbol\Phi_1\right\|^2+ \left\|\textbf{v}_2^H\textbf{H}_{rid}\textbf{A}\right\|^2+ \left\|\textbf{u}_{2}^H\text{diag}\{\textbf{h}_{id}^H\textbf{E}_{K}\}\right\|^2 + 1} \\
%&~~\text{s.t.}~~   |\boldsymbol\Theta_2(i,i)|=1,~~~\text{for}~i\in \cal E_L,  \label{givenA_Theta_2_1}\\
%&~~~~~~~|\boldsymbol\Theta_2(i,i)|=\beta_{2i},~\text{for}~i\in \cal E_K.  \label{givenA_Theta_2_2}
%\end{align}
%\end{subequations}
%where $\textbf{h}_{sir}=\textbf{A}(\textbf{h}_{sr} +\textbf{H}_{ir}\boldsymbol\Theta_1\textbf{h}_{si})$.
%In the same manner, the above problem can be translated to
\begin{subequations}\label{trV2}
\begin{align}
&~\max \limits_{\textbf{V}_2 } ~~~~~\frac{\text{tr}\{\textbf{H}_1\textbf{V}_2\}} {\text{tr}\{\textbf{H}_2\textbf{V}_2\}} \\
&~~\text{s.t.}~~~~~~~ \textbf{V}_2(i,i)=1, ~~\text{for}~i\in \cal E_L,\\
%&~~~~~~~~~~~~  \textbf{V}_2(i,i)>1, ~~\text{for}~i\in \cal E_K,\\
&~~~~~~~~~~~~~  \textbf{V}_2(N+1,N+1)=1, \\
&~~~~~~~~~~~~~  \text{tr}(\textbf{J}\textbf{V}_2)\leq\gamma_i,~\text{rank}(\textbf{V}_2)=1,~\textbf{V}_2\succeq\textbf{0},
\end{align}
\end{subequations}
where $\textbf{V}_2=\textbf{v}_2\textbf{v}_2^H$,
$\textbf{H}_1=\gamma_s\textbf{H}_{rid}\textbf{A}(\textbf{h}_{sr} +\textbf{H}_{ir}\boldsymbol\Theta_1\textbf{h}_{si}) [\textbf{H}_{rid}\textbf{A}(\textbf{h}_{sr} +\textbf{H}_{ir}\boldsymbol\Theta_1\textbf{h}_{si})]^H$,
$\textbf{H}_2=
\textbf{H}_{rid}\textbf{A}(\textbf{H}_{ir}\textbf{E}_{K}\boldsymbol\Theta_1\boldsymbol\Theta_1^H\textbf{E}_{K}\textbf{H}_{ir}^H+\textbf{I}_M)\textbf{A}^H\textbf{H}_{rid}^H +[\text{diag}\{\textbf{h}_{id}^H\textbf{E}_{K}\}\text{diag}\{ \textbf{E}_{K}\textbf{h}_{id}  \},~ \textbf{0}_{N \times 1};~
\textbf{0}_{1 \times N},~ 1]$ and $\textbf{J}=[\gamma_s\textbf{H}_3^H\textbf{H}_3+(\textbf{H}_4\textbf{H}_4^H+\textbf{H}_{ir}^H\textbf{A}\textbf{A}^H\textbf{H}_{ir}+\textbf{I}_N )\odot\textbf{E}_K,~ \textbf{0}_{N \times 1};~ \\
\textbf{0}_{1 \times N},~ 0]$,
wherein
$\textbf{H}_3=\textbf{E}_K\text{diag}\{ \textbf{H}_{ir}^T\textbf{A}^*(\textbf{h}_{sr}+\textbf{H}_{ir}\boldsymbol\Theta_1\textbf{h}_{si})^*\}$
and $\textbf{H}_4=\textbf{H}_{ir}^H\textbf{A}\textbf{H}_{ir}\textbf{E}_{K}\boldsymbol\Theta_1$.
Problem (\ref{trV2}) is a non-convex problem. After losing the rank constraint, it can be similarly translated to
\begin{subequations}\label{SDP_2}
\begin{align}
&\max \limits_{\widetilde{\textbf{V}}_2, \rho }  ~~~~\text{tr}(\textbf{H}_1\widetilde{\textbf{V}}_2) \\
&~\text{s.t.}~~~~~~ \widetilde{\textbf{V}}_2(i,i)=\rho, ~~\text{for}~i\in \cal E_L,\\
%&~~~~~~~~~~ \widetilde{\textbf{V}}_2(i,i)>\rho, ~~\text{for}~i\in \cal E_K,\\
&~~~~~~~~~~~ \widetilde{\textbf{V}}_2(N+1,N+1)=\rho,~\rho>0, \\
&~~~~~~~~~~~ \text{tr}(\textbf{J}\widetilde{\textbf{V}}_2)\leq\rho\gamma_i,~\text{tr}(\textbf{H}_2\widetilde{\textbf{V}}_2)=1,~ \widetilde{\textbf{V}}_2\succeq\textbf{0},
\end{align}
\end{subequations}
where $\rho=\text{tr}(\textbf{H}_2\textbf{V}_2)^{-1}$ is a slack variable and $\widetilde{\textbf{V}}_2=\rho\textbf{V}_2$. It is observed that (\ref{SDP_2}) is similar to (\ref{SDP}), thus (\ref{SDP_2}) can be solved in the same way as (\ref{SDP}). Finally the solutions $\textbf{v}_2$ and $\boldsymbol\Theta_2$ are obtained, and the details are omitted here for brevity.
The relationship between $\boldsymbol\Theta_2$ and $\textbf{v}_2$ is as follows
%\begin{equation}
%\boldsymbol\Theta_2=\text{diag}\left(\left[\frac{\textbf{v}_2}{\textbf{v}_2(N+1)}\right]^H(1:N)\right).
%\end{equation}
\begin{subnumcases}{\boldsymbol\Theta_2(i,i)=}{}
e^{j\text{arg}\left(\left[\frac{\textbf{v}_2(i)}{\textbf{v}_2(N+1)}\right]^*\right)},~~~~~~~~i\in \cal E_L, \\
\left[\frac{\textbf{v}_2(i)}{\textbf{v}_2(N+1)}\right]^*,~~~~~~~~~~i\in \cal E_K.
\end{subnumcases}

\subsection{Overall Algorithm and Complexity Analysis}
Since the objective function of problem (\ref{OP}) is non-decreasing and the transmit powers of S, AF relay and IRS active elements are limited, the objective function has an upper bound. Therefore, the convergence of the proposed HP-SDR-FP algorithm can be guaranteed. Our idea is alternative iteration, that is, the alternative iteration process are performed among $\textbf{A}$, $\boldsymbol\Theta_1$ and $\boldsymbol\Theta_2$ until the convergence criterion is satisfied, while the system rate is maximum. The proposed HP-SDR-FP method is summarized in Algorithm 1.
\begin{table}[h]\footnotesize
\renewcommand{\arraystretch}{1}
\centering
\begin{tabular}{p{440pt}}
\hline
$\bf{Algorithm~1}$  Proposed HP-SDR-FP Method\\
\hline
1.~ Initialize $\textbf{A}^0$, $\boldsymbol\Theta_1^0$ and $\boldsymbol\Theta_2^0$. According to (\ref{R}), $R^0$ can be obtained.\\
2.~ set the convergence error $\delta$ and the iteration number $t = 0$. \\
3.~\bf{repeat}  \\
4.~   Given $\boldsymbol\Theta_1^t$ and $\boldsymbol\Theta_2^t$, solve problem (\ref{t}) for $\widetilde{\textbf{A}}^{t+1}$, recover rank-1 solution $\widehat{\textbf{A}}^{t+1}$ via Gaussian randomization, obtain $\textbf{A}^{t+1}$.\\
5.~   Given $\textbf{A}^{t+1}$ and $\boldsymbol\Theta_2^t$, solve problem (\ref{SDP}) for $\widetilde{\textbf{V}}_1^{t+1}$, recover rank-1 solution $\textbf{V}_1^{t+1}$ via Gaussian randomization, obtain $\boldsymbol\Theta_1^{t+1}$.\\
6.   Given $\textbf{A}^{t+1}$ and $\boldsymbol\Theta_1^{t+1}$, solve problem (\ref{SDP_2}) for $\widetilde{\textbf{V}}_2^{t+1}$, recover rank-1 solution $\textbf{V}_2^{t+1}$ via Gaussian randomization, obtain $\boldsymbol\Theta_2^{t+1}$.\\
7.~   Calculate $R^{t + 1}$ by using $\textbf{A}^{t + 1}$, $\boldsymbol\Theta_1^{t + 1}$ and $\boldsymbol\Theta_2^{t + 1}$.\\
8.~   Update $t = t + 1$.\\
9.~\bf{until} \\
~~~    $\left| {R^{t+1}- R^t} \right|\le \delta$.\\
\hline
\end{tabular}
\end{table}
%\begin{table*}[ht]\normalsize %hb代表放在文章底部，%ht为放在文章顶部
% \caption{XXX}
%   \centering
%    \begin{tabular}{|c|c|c|c|}
%     	\hline
%         Problem & The number of linear constraints & The size of one LMI constraint & The number of decision variables \\
%         \hline
%          (\ref{t})      &  4         &  $M^2$      & $n_\textbf{A}=M^4+1$          \\
%          \hline
%          (\ref{SDP})    &  $N+6$     &   $N+1$     & $n_{\textbf{V}_1}=(N+1)^2+1$      \\
%          \hline
%          (\ref{SDP_2})  &  $N+4$     &   $N+1$     & $n_{\textbf{V}_2}=(N+1)^2+1$      \\
%          \hline   	
%    \end{tabular}
%\end{table*}

After that, the complexity of Algorithm 1 is calculated and analyzed according to problems (\ref{t}), (\ref{SDP}) and (\ref{SDP_2}). Problem (\ref{t}) has $4$ linear constraints with dimension 1, one linear matrix inequality (LMI) constraint of size $M^2$ and $M^4+1$ decision variables. Hence, the computational complexity of problem (\ref{t}) is denoted as
\begin{equation}
\mathcal{O}\{n_\textbf{A}\sqrt{M^2+4}(M^6+4 + n_\textbf{A}(M^4+4) + n_\textbf{A}^2)\text{ln}(1/\varepsilon)\}
\end{equation}
float-point operations (FLOPs), where $n_\textbf{A}=M^4+1$ and $\varepsilon$ represents the computation accuracy. For problem (\ref{SDP}), there exit $L+6$ linear constraints with dimension 1, one LMI constraint of size $N+1$ and $(N + 1)^2+ 1$ decision variables, so the computational complexity of problem (\ref{SDP}) is given by
\begin{equation}
\mathcal{O}\{n_{\textbf{V}_1}\sqrt{N+L+7}((N+1)^3+L+6 + n_{\textbf{V}_1}((N+1)^2+L+6) + n_{\textbf{V}_1}^2)\text{ln}(1/\varepsilon)\}
\end{equation}
FLOPs, where $n_{\textbf{V}_1}=(N + 1)^2+ 1$. For problem (\ref{SDP_2}), there are $L+4$ linear constraints with dimension 1, one LMI constraint of size $N+1$ and $(N + 1)^2+ 1$ decision variables. Thus, the computational complexity of problem (\ref{SDP_2}) is expressed as
\begin{equation}
\mathcal{O}\{n_{\textbf{V}_2}\sqrt{N+L+5}((N+1)^3+L+4 + n_{\textbf{V}_2}((N+1)^2+L+4) + n_{\textbf{V}_2}^2)\text{ln}(1/\varepsilon)\}
\end{equation}
FLOPs, where $n_{\textbf{V}_2}=(N + 1)^2+ 1$. Therefore, the total computational complexity of Algorithm 1 is written by
\begin{align}
%D(\mathcal{O}_\textbf{A}+\mathcal{O}_{\textbf{V}_1}+\mathcal{O}_{\textbf{V}_2})
&\mathcal{O}\{D_1[n_\textbf{A}\sqrt{M^2+4}(M^6+4+n_\textbf{A}(M^4+4)+n_\textbf{A}^2)  + n_{\textbf{V}_1}\sqrt{N+L+7}((N+1)^3+L+6+n_{\textbf{V}_1}((N+1)^2  \nonumber\\
&+L+6) + n_{\textbf{V}_1}^2)+n_{\textbf{V}_2}\sqrt{N+L+5}((N+1)^3+L+4  +n_{\textbf{V}_2}((N+1)^2+L+4)+n_{\textbf{V}_2}^2)]\text{ln}(1/\varepsilon)\}
\end{align}
FLOPs, where $D_1$ is the maximum number of alternating iterations needed for convergence in Algorithm 1. It is obvious that the highest order of computational complexity is $M^{13}$ and $N^{6.5}$ FLOPs.

\section{Proposed a Low-complexity WF-GPI-GRR-based Max-SNR Method}\label{WF_GPI_GRR}
In what follows, to reduce the computational complexity, a low-complexity WF-GPI-GRR-based Max-SNR method is put forward. For gaining rate enhancement, we apply WF operation to exploit the colored property of noise and present the related system model. Here, active IRS reflecting coefficient matrix is split into amplifying coefficient and IRS phase-shift matrix. The details of derivation on the amplifying coefficients, AF relay beamforming matrix and IRS phase-shift matrices are described as below.

\subsection{System Model}
For brevity, it is assumed that the amplifying coefficients of each IRS active element in the first time slot and the second time slot are $|\beta_1|$ and $|\beta_2|$, respectively. Let us define
\begin{subequations}
\begin{align}
&\boldsymbol\Psi_1=\overline{\textbf{E}}_K\widehat{\boldsymbol\Theta}_1,~\boldsymbol\Phi_1=|\beta_1|\textbf{E}_K\widehat{\boldsymbol\Theta}_1, \\
&\boldsymbol\Psi_2=\overline{\textbf{E}}_K\widehat{\boldsymbol\Theta}_2,~\boldsymbol\Phi_2=|\beta_2|\textbf{E}_K\widehat{\boldsymbol\Theta}_2,
\end{align}
\end{subequations}
where the phase-shift matrix $\widehat{\boldsymbol\Theta}_1 = \text{diag}( e^{j\theta _{1i}}, \cdots,e^{j\theta _{1N}} )$, $\widehat{\boldsymbol\Theta}_2 = \text{diag}( e^{j\theta _{2i}}, \cdots,e^{j\theta _{2N}} )$, $|\widehat{\boldsymbol\Theta}_1(i,i)|=1$ and $|\widehat{\boldsymbol\Theta}_2(i,i)|=1$. Thus we have
\begin{equation}\label{redefine_Theta}
\boldsymbol\Theta_1=(\overline{\textbf{E}}_K+|\beta_1|\textbf{E}_K)\widehat{\boldsymbol\Theta}_1,~
\boldsymbol\Theta_2=(\overline{\textbf{E}}_K+|\beta_2|\textbf{E}_K)\widehat{\boldsymbol\Theta}_2.
\end{equation}
In the first time slot, the received signal at AF relay can be redescribed as
\begin{equation}\label{y_r_redes}
\textbf{y}_r=\sqrt {P_s}[\textbf{h}_{sr} + \textbf{H}_{ir}( \overline{\textbf{E}}_K+|\beta_1|\textbf{E}_K )\widehat{\boldsymbol\Theta}_1\textbf{h}_{si}]x + \underbrace{(|\beta_1|\textbf{H}_{ir}\textbf{E}_K\widehat{\boldsymbol\Theta}_1\textbf{n}_{1i} + \textbf{n}_r)}_{\textbf{n}_{1r}}.
\end{equation}
As matter of fact, $\textbf{n}_{1r}$ is color, not white. It is necessary for us to whiten the color noise $\textbf{n}_{1r}$ by using covariance matrix $\textbf{C}_n$. The covariance $W_{1r}$ of $\textbf{n}_{1r}$ is given by
\begin{equation}
W_{1r}=\beta_1^2\| \textbf{H}_{ir}\textbf{E}_K\widehat{\boldsymbol\Theta}_1 \|_F^2\sigma^2 + \sigma^2.
\end{equation}
While $\textbf{n}_{1i}$ and $\textbf{n}_r$ are the independent and identically distributed random vectors, $\textbf{n}_{1r}$ has a mean vector of all-zeros and covariance matrix
\begin{equation}
\textbf{C}_{1r}=\beta_1^2\sigma^2\textbf{H}_{ir}\textbf{E}_K\widehat{\boldsymbol\Theta}_1\widehat{\boldsymbol\Theta}_1^H\textbf{E}_K\textbf{H}_{ir}^H + \sigma^2{\bf I}_{M},
\end{equation}
where obviously $\textbf{C}_{1r}$ is a positive definite matrix. Defining the WF matrix $\textbf{W}_{1r}$ with $\textbf{W}_{1r}\textbf{W}_{1r}^H = \textbf{C}_{1r}^{-1}$, which yields
\begin{equation}
\textbf{W}_{1r} = \textbf{C}_{1r}^{-\frac{1}{2}} = (\textbf{Q}_{1r}\mathbf{\Lambda}_{1r}\textbf{Q}_{1r}^H)^{-\frac{1}{2}}=\textbf{Q}_{1r}\mathbf{\Lambda}_{1r}^{-\frac{1}{2}}\textbf{Q}_{1r}^H,
\end{equation}
where $\textbf{Q}_{1r}$ is an unitary matrix, and $\mathbf{\Lambda}_{1r}$ is a diagonal matrix consisting of eigenvalues. Performing the WF operation to (\ref{y_r_redes}) yields
\begin{equation}
\overline{\textbf{y}}_r=\sqrt {P_s}\textbf{W}_{1r}[\textbf{h}_{sr} + \textbf{H}_{ir}( \overline{\textbf{E}}_K+|\beta_1|\textbf{E}_K )\widehat{\boldsymbol\Theta}_1\textbf{h}_{si}]x + \underbrace{\textbf{W}_{1r}(|\beta_1|\textbf{H}_{ir}\textbf{E}_K\widehat{\boldsymbol\Theta}_1\textbf{n}_{1i} + \textbf{n}_r)}_{\overline{\textbf{n}}_{1r}},
\end{equation}
where $\overline{\textbf{n}}_{1r}$ is the standard white noise with covariance matrix ${\bf I}_{M}$. The transmit signal at AF relay is $\overline{\textbf{y}}_t=\textbf{A}\overline{\textbf{y}}_r$. In the second time slot, the received signal at D is denoted as
\begin{align}
&y_d= \sqrt {P_s}[\textbf{h}_{rd}^H + \textbf{h}_{id}^H(\overline{\textbf{E}}_K+|\beta_2|\textbf{E}_K)\widehat{\boldsymbol\Theta}_2\textbf{H}_{ir}^H]\textbf{A}\textbf{W}_{1r} [\textbf{h}_{sr} +\textbf{H}_{ir}(\overline{\textbf{E}}_K+|\beta_1|\textbf{E}_K)\widehat{\boldsymbol\Theta}_1\textbf{h}_{si}]x \nonumber\\
&~~~~~~~ + [\textbf{h}_{rd}^H + \textbf{h}_{id}^H(\overline{\textbf{E}}_K+|\beta_2|\textbf{E}_K)\widehat{\boldsymbol\Theta}_2\textbf{H}_{ir}^H]\textbf{A}\overline{\textbf{n}}_{1r} + |\beta_2|\textbf{h}_{id}^H\textbf{E}_K\widehat{\boldsymbol\Theta}_2\textbf{n}_{2i} + \text{n}_d.
\end{align}
The corresponding SNR can be represented as
\begin{equation}\label{SNR_3}
\text{SNR}=\frac{P_d}{N_d},
\end{equation}
where
$P_d$ is the received signal power at D, and
$P_d=\gamma_s|[\textbf{h}_{rd}^H + \textbf{h}_{id}^H(\overline{\textbf{E}}_K+|\beta_2|\textbf{E}_K)\widehat{\boldsymbol\Theta}_2\textbf{H}_{ir}^H]\textbf{A}\textbf{W}_{1r}
[\textbf{h}_{sr} + \textbf{H}_{ir}(\overline{\textbf{E}}_K+|\beta_1|\textbf{E}_K)\widehat{\boldsymbol\Theta}_1\textbf{h}_{si}]|^2$.
$N_d$ is the received noise power at D, and
$N_d=\beta_1^2\|[\textbf{h}_{rd}^H + \textbf{h}_{id}^H(\overline{\textbf{E}}_K+|\beta_2|\textbf{E}_K)\widehat{\boldsymbol\Theta}_2\textbf{H}_{ir}^H]
\textbf{A}\textbf{W}_{1r}\textbf{H}_{ir}\textbf{E}_K\widehat{\boldsymbol\Theta}_1\|^2+ \|[\textbf{h}_{rd}^H + \textbf{h}_{id}^H(\overline{\textbf{E}}_K+|\beta_2|\textbf{E}_K)\widehat{\boldsymbol\Theta}_2\textbf{H}_{ir}^H]\textbf{A}\textbf{W}_{1r}\|^2+ \beta_2^2\|\textbf{h}_{id}^H\textbf{E}_K\widehat{\boldsymbol\Theta}_2\|^2 + 1$.
%\begin{small}
%\begin{align}\label{SNR_3}
%&\text{SNR}= \nonumber\\
%&\frac{\gamma_s|[\textbf{h}_{rd}^H + \textbf{h}_{id}^H(\overline{\textbf{E}}_K+|\beta_2|\textbf{E}_K)\widehat{\boldsymbol\Theta}_2\textbf{H}_{ir}^H]\textbf{A}\textbf{W}_{1r}
%[\textbf{h}_{sr} + \textbf{H}_{ir}(\overline{\textbf{E}}_K+|\beta_1|\textbf{E}_K)\widehat{\boldsymbol\Theta}_1\textbf{h}_{si}]|^2}  {\beta_1^2\|[\textbf{h}_{rd}^H + \textbf{h}_{id}^H(\overline{\textbf{E}}_K+|\beta_2|\textbf{E}_K)\widehat{\boldsymbol\Theta}_2\textbf{H}_{ir}^H]
%\textbf{A}\textbf{W}_{1r}\textbf{H}_{ir}\textbf{E}_K\widehat{\boldsymbol\Theta}_1\|^2+ \|[\textbf{h}_{rd}^H + \textbf{h}_{id}^H(\overline{\textbf{E}}_K+|\beta_2|\textbf{E}_K)\widehat{\boldsymbol\Theta}_2\textbf{H}_{ir}^H]\textbf{A}\textbf{W}_{1r}\|^2+ \beta_2^2\|\textbf{h}_{id}^H\textbf{E}_K\widehat{\boldsymbol\Theta}_2\|^2 + 1}.
%\end{align}
%\end{small}
It is assumed that the power budgets $P_s$, $P_r$ and $P_i$ are respectively fully used to transmit signals at S, AF relay and IRS. Therefore, the optimization problem can be converted to
\begin{subequations}
\begin{align}
&\max \limits_{|\beta_1|, |\beta_2|, \widehat{\boldsymbol\Theta}_1, \widehat{\boldsymbol\Theta}_2, \textbf{A} }~~~~ \text{(\ref{SNR_3})} \\
&~~~~~~~~\text{s.t.}~~~~~~~~~~~~ |\widehat{\boldsymbol\Theta}_1(i,i)|=1,~~~|\widehat{\boldsymbol\Theta}_2(i,i)|=1.
\end{align}
\end{subequations}
It is necessary to solve the above problem for optimal $|\beta_1|$, $|\beta_2|$, $\widehat{\boldsymbol\Theta}_1$, $\widehat{\boldsymbol\Theta}_2$ and $\textbf{A}$.

\subsection{Solve $|\beta_1|$ and $|\beta_2|$}
In the first time slot, the reflected signal at IRS is written by
\begin{equation}\label{y_1i_t}
\textbf{y}_{1i}^t=\sqrt {{P_s}}\boldsymbol\Theta_1\textbf{h}_{si}x+\boldsymbol\Phi_1\textbf{n}_{1i}
=\underbrace{\sqrt {{P_s}}\overline{\textbf{E}}_K\widehat{\boldsymbol\Theta}_1\textbf{h}_{si}x}_{\textbf{y}_{1i}^{pt}}
+\underbrace{\sqrt {{P_s}}|\beta_1|\textbf{E}_K\widehat{\boldsymbol\Theta}_1\textbf{h}_{si}x+|\beta_1|\textbf{E}_K\widehat{\boldsymbol\Theta}_1\textbf{n}_{1i}}_{\textbf{y}_{1i}^{at}},
\end{equation}
where $\textbf{y}_{1i}^{pt}$ and $\textbf{y}_{1i}^{at}$ are respectively the signals reflected by passive elements $\cal E_L$ and active elements $\cal E_K$.  Additionally, the power consumed by the active elements is $P_i$. We have
\begin{align}\label{P_i_1}
&P_i=
P_s\beta_1^2 \|\textbf{E}_K\widehat{\boldsymbol\Theta}_1\textbf{h}_{si}\|^2+\beta_1^2\|\textbf{E}_K\widehat{\boldsymbol\Theta}_1\|_F^2\sigma_{1i}^2  \\ &~~~=\beta_1^2P_s\sum_{k=1}^K| e^{j\theta _{1k}}h_{si}^k  |^2 +\beta_1^2\sum_{k=1}^K| e^{j\theta _{1k}}|^2\sigma_{1i}^2 \nonumber\\
&~~~=\beta_1^2P_s\sum_{k=1}^K| h_{si}^k |^2 +K\beta_1^2\sigma^2,
\end{align}
where $\theta _{1k}$ is the phase shift of the $k$th IRS active element in the first time slot, $h_{si}^k$ is the channel between S and the $k$th IRS active element and follows Rayleigh distribution with the following expression
\begin{equation}
h_{si}^k=\sqrt{PL_{si}^k}g_{si}^ke^{-j\varphi _{sk}},
\end{equation}
where $PL_{si}^k$, $g_{si}^k$ and $\varphi _{sk}$ denote the path loss, the channel gain and the channel phase from S to the $k$th IRS active element, respectively. $| g_{si}^k |^2$ follows Exponential distribution  \cite{2022MN}, and the corresponding probability density function is given by
\begin{subnumcases}{f_{| g_{si}^k |^2}(x)=}{}
\frac{1}{\lambda_{si}} e^{-\frac{x}{\lambda_{si}}},~~~~~~  x\in [0,+\infty) \\
0,~~~~~~~~~~~~~~~~~  \text{otherwise}
\end{subnumcases}
where $\lambda_{si}$ is the Exponential distribution parameter. Let us define $PL_{si}^k$ is equal to the path loss from S to IRS (i.e., $PL_{si}$). Using the weak law of large numbers, (\ref{P_i_1}) can be further written as
\begin{align}
&P_i=\beta_1^2P_s\sum_{k=1}^K| \sqrt{PL_{si}}g_{si}^ke^{-j\varphi _{sk}}  |^2 +K\beta_1^2\sigma^2 \nonumber\\
&~~~=\beta_1^2P_sPL_{si}\sum_{k=1}^K| g_{si}^k |^2 +K\beta_1^2\sigma^2  \nonumber\\
&~~~\approx K\beta_1^2P_sPL_{si}\cdot\mathbb{E}(| g_{si}^k |^2)+K\beta_1^2\sigma^2  \nonumber\\
&~~~=K\beta_1^2P_sPL_{si}\lambda_{si}+K\beta_1^2\sigma^2,
\end{align}
$\beta_1$ can be achieved as
\begin{equation}\label{beta1}
|\beta_1|=\sqrt{\frac{P_i}{KP_sPL_{si}\lambda_{si} + K\sigma^2}}.
\end{equation}
Similarly, the received signal of the $k$th active IRS element in the second time slot is
\begin{equation}
y_{2i}^{rk}=\textbf{h}_{rk}^H\textbf{A}\overline{\textbf{y}}_r+\text{n}_{2i,k}=\textbf{h}_{rk}^H\overline{\textbf{y}}_t+\text{n}_{2i,k},
\end{equation}
where $\textbf{h}_{rk}^H\in \mathbb C^{1 \times M}$ represents the channel between AF relay and the $k$th active IRS element.
\begin{equation}
\textbf{h}_{rk}^H=[\sqrt{PL_{ri}^{1k}}g_{ri}^{1k}e^{-j\varphi _{1k}},\cdots, \sqrt{PL_{ri}^{Mk}}g_{ri}^{Mk}e^{-j\varphi _{Mk}}],
\end{equation}
where $PL_{ri}^{mk}$, $g_{ri}^{mk}$ and $\varphi _{mk}$ are the path loss, the channel gain and the channel phase between the $m$th antenna at AF relay and the $k$th active IRS element. Defining $PL_{ri}^{mk}=PL_{ri}$, $PL_{ri}$ is the path loss from AF relay to IRS.
The reflected signal of the $k$th active IRS element is
\begin{align}
&y_{2i}^{tk}=|\beta_2|e^{j\theta _{2k}}\textbf{h}_{rk}^H\overline{\textbf{y}}_t+|\beta_2|e^{j\theta _{2k}}\text{n}_{2i,k} \nonumber\\
&~~~~=|\beta_2|| \textbf{h}_{rk}^H\overline{\textbf{y}}_t |e^{j({\theta _{2k}+\varphi _{rkt}})}+|\beta_2|e^{j\theta _{2k}}\text{n}_{2i,k}  \nonumber\\
&~~~~=|\beta_2|| \textbf{h}_{rk}^H | | \overline{\textbf{y}}_t |e^{j(\theta _{2k}+\varphi _{rkt})}+|\beta_2|e^{j\theta _{2k}}\text{n}_{2i,k},
\end{align}
where $\theta _{2k}$ is the phase shift of the $k$th IRS active element in the second time slot, $\varphi _{rkt}$ is the phase of $\textbf{h}_{rk}^H\overline{\textbf{y}}_t$.
It is assumed that the transmit power of AF relay is $P_r$, the corresponding power of the reflected signal of the $k$th active IRS element is
\begin{align}
&P_{2i}^{tk}=\beta_2^2| \textbf{h}_{rk}^H |^2 | \overline{\textbf{y}}_t |^2+\beta_2^2\sigma_{2i,k}^2  \nonumber\\
&~~~~~=\beta_2^2P_rPL_{ri}\sum_{m=1}^M| g_{ri}^{mk}|^2 + \frac{\beta_2^2\sigma^2}{K}   \nonumber\\
&~~~~~\approx M\beta_2^2P_rPL_{ri}\cdot\mathbb{E}(| g_{ri}^{mk} |^2)+\frac{\beta_2^2\sigma^2}{K}  \nonumber\\
&~~~~~=M\beta_2^2P_rPL_{ri}\lambda_{ri}+\frac{\beta_2^2\sigma^2}{K},
\end{align}
where $\sigma_{2i,k}^2=\sigma_{2i}^2/K=\sigma^2/K$, $\lambda_{ri}$ is the Exponential distribution parameter of channel
from AF relay to IRS. Thus the power of the reflected signal of $K$ active IRS elements is
\begin{equation}
P_i=\sum_{k=1}^KP_{2i}^{tk}=KM\beta_2^2P_rPL_{ri}\lambda_{ri}+\beta_2^2\sigma^2,
\end{equation}
which yields
\begin{equation}\label{beta2}
|\beta_2|=\sqrt{\frac{P_i}{KMP_rPL_{ri}\lambda_{ri}+\sigma^2}}.
\end{equation}

\subsection{Optimize $\textbf{A}$ Given $\widehat{\bf\Theta}_1$ and $\widehat{\bf\Theta}_2$}
Aiming at maximizing the received signal power, MRC-MRT method are applied to solve $\textbf{A}$ as follows
\begin{align}\label{A1}
\textbf{A}&=A\frac{[\textbf{h}_{rd}+ \textbf{H}_{ir}\widehat{\boldsymbol\Theta}_2^H(\overline{\textbf{E}}_K+|\beta_2|\textbf{E}_K)\textbf{h}_{id}]}{\|\textbf{h}_{rd}^H + \textbf{h}_{id}^H(\overline{\textbf{E}}_K+|\beta_2|\textbf{E}_K)\widehat{\boldsymbol\Theta}_2\textbf{H}_{ir}^H\|} \frac{[\textbf{h}_{sr} +
\textbf{H}_{ir}(\overline{\textbf{E}}_K+|\beta_1|\textbf{E}_K)\widehat{\boldsymbol\Theta}_1\textbf{h}_{si}]^H\textbf{W}_{1r}^H}{\|\textbf{W}_{1r}[\textbf{h}_{sr} +\textbf{H}_{ir}(\overline{\textbf{E}}_K+|\beta_1|\textbf{E}_K)\widehat{\boldsymbol\Theta}_1\textbf{h}_{si}]\|} \nonumber\\
&=A\boldsymbol\Upsilon,
\end{align}
where $A$ is the amplify factor of AF relay. Since the transmit power of AF relay is $P_r$, we have
\begin{equation}\label{A2}
A = \sqrt{\frac{\gamma_r}{\gamma_s\|\boldsymbol\Upsilon\textbf{W}_{1r}[\textbf{h}_{sr} + \textbf{H}_{ir}\boldsymbol(\overline{\textbf{E}}_K+|\beta_1|\textbf{E}_K)\widehat{\boldsymbol\Theta}_1\textbf{h}_{si}]\|^2   + \beta_1^2\|\boldsymbol\Upsilon\textbf{W}_{1r}\textbf{H}_{ir}\textbf{E}_K\widehat{\boldsymbol\Theta}_1\|_F^2+\|\boldsymbol\Upsilon\textbf{W}_{1r}\|_F^2}}.
\end{equation}
Inserting $A$ back into (\ref{A1}), $\textbf{A}$ can be obtained.

\subsection{Optimize $\widehat{\bf\Theta}_1$ Given $\textbf{A}$ and $\widehat{\bf\Theta}_2$}
By defining $\widehat{\textbf{u}}_1= [ e^{j\theta _{1i}}, \cdots,e^{j\theta _{1N}} ]^T$, $\widehat{\textbf{v}}_1=[ \widehat{\textbf{u}}_1; 1]$ and $\widehat{\textbf{H}}_{sir}=[\textbf{H}_{ir}(\overline{\textbf{E}}_K+|\beta_1|\textbf{E}_K)\text{diag}\{\textbf{h}_{si}\}, \textbf{h}_{sr}]$. Given AF relay beamforming matrix $\textbf{A}$ and $\widehat{\boldsymbol\Theta}_2$, the optimization problem is equivalent to
\begin{subequations}
\begin{align}
&\max \limits_{\widehat{\textbf{v}}_1 } ~~~\frac{\widehat{\textbf{v}}_1^H\widehat{\textbf{F}}_1\widehat{\textbf{v}}_1}{\widehat{\textbf{v}}_1^H\widehat{\textbf{F}}_2\widehat{\textbf{v}}_1 } \\
&~\text{s.t.}~~~~~  |\widehat{\textbf{v}}_1(i)|=1,~\forall i=1,2,\cdots,N, \\
&~~~~~~~~~~ \widehat{\textbf{v}}_1(N+1)=1,
\end{align}
\end{subequations}
where $\widehat{\textbf{F}}_1$ is a Hermitian matrix and
$\widehat{\textbf{F}}_1=\gamma_s\widehat{\textbf{H}}_{sir}^H[(\textbf{h}_{rd}^H + \textbf{h}_{id}^H\boldsymbol(\overline{\textbf{E}}_K+|\beta_2|\textbf{E}_K)\widehat{\boldsymbol\Theta}_2\textbf{H}_{ir}^H)
\textbf{A}\textbf{W}_{1r}]^H[(\textbf{h}_{rd}^H + \textbf{h}_{id}^H\boldsymbol(\overline{\textbf{E}}_K+|\beta_2|\textbf{E}_K)
\widehat{\boldsymbol\Theta}_2\textbf{H}_{ir}^H)\textbf{A}\cdot\\ \textbf{W}_{1r}]\widehat{\textbf{H}}_{sir}$.
%where $\widehat{\textbf{h}}_{rid}=[(\textbf{h}_{rd}^H + \textbf{h}_{id}^H\boldsymbol(\overline{\textbf{E}}_K+|\beta_2|\textbf{E}_K)\widehat{\boldsymbol\Theta}_2
%\textbf{H}_{ir}^H)\textbf{A}\textbf{W}_{1r}]^H$.
$\widehat{\textbf{F}}_2$ is a positively semi-definite Hermitian matrix and
$\widehat{\textbf{F}}_2=[
\beta_1^2\text{diag}\{\textbf{E}_K\textbf{H}_{ir}^H\widehat{\textbf{h}}_{rid}\}\text{diag}\{\widehat{\textbf{h}}_{rid}^H\textbf{H}_{ir}\textbf{E}_K\},~ \textbf{0}_{N \times 1};~
\textbf{0}_{1 \times N},~ \\ \|\widehat{\textbf{h}}_{rid}^H\|^2+ \beta_2^2\|\textbf{h}_{id}^H\textbf{E}_K\widehat{\boldsymbol\Theta}_2\|^2 + 1]$.
The above problem can be relaxed to
\begin{subequations}
\begin{align}
&\max \limits_{\widehat{\textbf{v}}_1 } ~~~\frac{\widehat{\textbf{v}}_1^H\widehat{\textbf{F}}_1\widehat{\textbf{v}}_1}{\widehat{\textbf{v}}_1^H\widehat{\textbf{F}}_2\widehat{\textbf{v}}_1 } \\
&~\text{s.t.}~~~~  \|\widehat{\textbf{v}}_1\|^2=N+1,
\end{align}
\end{subequations}
which can be constructed as
\begin{subequations}\label{v1_hat}
\begin{align}
&\max \limits_{\widehat{\textbf{v}}_1 } ~~~\frac{\widehat{\textbf{v}}_1^H\widehat{\textbf{F}}_1\widehat{\textbf{v}}_1}{\widehat{\textbf{v}}_1^H\widehat{\textbf{F}}_2\widehat{\textbf{v}}_1 }\cdot\frac{\widehat{\textbf{v}}_1^H\textbf{I}_{N+1}\widehat{\textbf{v}}_1}{\widehat{\textbf{v}}_1^H\textbf{I}_{N+1}\widehat{\textbf{v}}_1 } \\
&~\text{s.t.}~~~~  \|\widehat{\textbf{v}}_1\|^2=N+1.
\end{align}
\end{subequations}
$\widehat{\textbf{v}}_1$ can be solved by using GPI algorithm, the details of GPI procedure is presented in Algorithm 2,
where we define $\boldsymbol\Omega(\widehat{\textbf{v}}_1^t)=(\widehat{\textbf{v}}_1^H\widehat{\textbf{F}}_1\widehat{\textbf{v}}_1)\textbf{I}_{N+1}+
(\widehat{\textbf{v}}_1^H\textbf{I}_{N+1}\widehat{\textbf{v}}_1)\widehat{\textbf{F}}_1$ and $\boldsymbol\Xi_1(\widehat{\textbf{v}}_1^t)=(\widehat{\textbf{v}}_1^H\widehat{\textbf{F}}_2\widehat{\textbf{v}}_1)\textbf{I}_{N+1}+
(\widehat{\textbf{v}}_1^H\textbf{I}_{N+1}\widehat{\textbf{v}}_1)\widehat{\textbf{F}}_2$.
\begin{table}[h]\footnotesize
\renewcommand{\arraystretch}{1}
\centering
\begin{tabular}{p{440pt}}
\hline
$\bf{Algorithm~2}$  GPI Algorithm to Compute Phase-Shift Vector $\widehat{\textbf{v}}_1$ with Given $\textbf{A}$ and $\widehat{\boldsymbol\Theta}_2$\\
\hline
1.~~ Given $\textbf{A}$ and $\widehat{\boldsymbol\Theta}_2$, and initialize $\widehat{\textbf{v}}_1^0$.\\
2.~~ Set the tolerance factor $\xi$ and the iteration number $t = 0$. \\
3.~\bf{repeat}  \\
4.~~   Compute the function matrix $\boldsymbol\Omega(\widehat{\textbf{v}}_1^t)$ and $\boldsymbol\Xi_1(\widehat{\textbf{v}}_1^t)$.\\
5.~~   Calculate $\textbf{y}^t= \boldsymbol\Xi_1(\widehat{\textbf{v}}_1^t)^\dag\boldsymbol\Omega(\widehat{\textbf{v}}_1^t)\widehat{\textbf{v}}_1^t$. \\
6.~~   Update $\widehat{\textbf{v}}_1^{t+1}=\frac{\textbf{y}^t}{\|\textbf{y}^t\|}$.\\
7.~~   Update $t = t + 1$.\\
8.~\bf{until} \\
~~~~    $\| {\widehat{\textbf{v}}_1^{t+1}- \widehat{\textbf{v}}_1^t} \|\le \xi$.\\
\hline
\end{tabular}
\end{table}

\subsection{Optimize $\widehat{\bf\Theta}_2$ Given $\textbf{A}$ and $\widehat{\bf\Theta}_1$}
If $\textbf{A}$ and $\widehat{\boldsymbol\Theta}_1$ are fixed, let us define $\widehat{\textbf{u}}_2= [ e^{j\theta _{2i}}, \cdots,e^{j\theta _{2N}} ]^H$, $\widehat{\textbf{v}}_2=[ \widehat{\textbf{u}}_2; 1]$ and  $\widehat{\textbf{H}}_{rid}=[\text{diag}\{\textbf{h}_{id}^H(\overline{\textbf{E}}_K+|\beta_2|\textbf{E}_K)\}\textbf{H}_{ir}^H; \textbf{h}_{rd}^H]$. Accordingly, the optimization problem is reduced to
\begin{subequations}
\begin{align}
&\max \limits_{\widehat{\textbf{v}}_2 } ~~~\frac{\widehat{\textbf{v}}_2^H\widehat{\textbf{H}}_1\widehat{\textbf{v}}_2}{\widehat{\textbf{v}}_2^H\widehat{\textbf{H}}_2\widehat{\textbf{v}}_2 } \\
&~\text{s.t.}~~~~  |\widehat{\textbf{v}}_2(i)|=1,~\forall i=1,2,\cdots,N, \\
&~~~~~~~~~ \widehat{\textbf{v}}_2(N+1)=1,
\end{align}
\end{subequations}
where $\widehat{\textbf{H}}_1$ is a Hermitian matrix and
$\widehat{\textbf{H}}_1=
\gamma_s\widehat{\textbf{H}}_{rid}\textbf{A}\textbf{W}_{1r}[\textbf{h}_{sr} +\textbf{H}_{ir}(\overline{\textbf{E}}_K+|\beta_1|\textbf{E}_K)\widehat{\boldsymbol\Theta}_1\textbf{h}_{si}] \{\widehat{\textbf{H}}_{rid}\textbf{A}\textbf{W}_{1r}[\textbf{h}_{sr} +\textbf{H}_{ir}(\overline{\textbf{E}}_K+|\beta_1|\textbf{E}_K)\widehat{\boldsymbol\Theta}_1\textbf{h}_{si}]\}^H$. $\widehat{\textbf{H}}_2$ is a positively definite Hermitian matrix and
$\widehat{\textbf{H}}_2=
\widehat{\textbf{H}}_{rid}\textbf{A}\textbf{W}_{1r}(\beta^2_1\textbf{H}_{ir}\textbf{E}_K\\ \widehat{\boldsymbol\Theta}_1
\widehat{\boldsymbol\Theta}_1^H\textbf{E}_K
\textbf{H}_{ir}^H+\textbf{I}_M)\textbf{W}_{1r}^H\textbf{A}^H\widehat{\textbf{H}}_{rid}^H+\left[
\begin{array}{cccc}
\beta^2_2\text{diag}\{\textbf{h}_{id}^H\textbf{E}_{K}\}\text{diag}\{ \textbf{E}_{K}\textbf{h}_{id} \},~ \textbf{0}_{N \times 1};~
\textbf{0}_{1 \times N}, & 1
\end{array}
\right]$.
We have the following relaxed transformation
\begin{subequations}\label{v2_hat}
\begin{align}
&\max \limits_{\widetilde{\textbf{v}}_2 } ~~~\frac{\widetilde{\textbf{v}}_2^H\widehat{\textbf{H}}_1\widetilde{\textbf{v}}_2}{\widetilde{\textbf{v}}_2^H\widehat{\textbf{H}}_2\widetilde{\textbf{v}}_2 } \\
&~\text{s.t.}~~~~~   \widetilde{\textbf{v}}_2^H\widetilde{\textbf{v}}_2=1
\end{align}
\end{subequations}
where $\widetilde{\textbf{v}}_2=\frac{\widehat{\textbf{v}}_2}{\sqrt{N+1}}$. Moreover, in line with the GRR theorem, the optimal $\widetilde{\textbf{v}}_2$ is obtained as the eigenvector corresponding to the largest eigenvalue of $\widehat{\textbf{H}}_2^{-1}\widehat{\textbf{H}}_1$. Thereby $\widehat{\textbf{v}}_2$ and $\widehat{\boldsymbol\Theta}_2$ is achieved.

\subsection{Overall Algorithm and Complexity Analysis}
The proposed the low-complexity WF-GPI-GRR method is summarized in Algorithm 3. The main idea consists of two parts: the amplifying coefficient of active element and the iterative idea. The analytic solutions of amplifying coefficients of IRS active elements in the first time slot and the second time slot, i.e., $\beta_1$ and $\beta_2$, are determined by the transmit power of S, AF relay, IRS. Furthermore, $\beta_1$ and $\beta_2$ are denoted as (\ref{beta1}) and (\ref{beta2}). The iterative idea can be described as follows: for given $\widehat{\boldsymbol\Theta}_1$ and $\widehat{\boldsymbol\Theta}_2$, the closed-form expression of $\textbf{A}$ are represented as (\ref{A1}) by utilizing MRC-MRT; for given $\textbf{A}$ and $\widehat{\boldsymbol\Theta}_2$, GPI is applied to achieve $\widehat{\boldsymbol\Theta}_1$; for given $\textbf{A}$ and $\widehat{\boldsymbol\Theta}_1$, $\widehat{\boldsymbol\Theta}_2$ is obtained in a closed-form expression by using GRR theorem. The alternative iteration process are performed among $\textbf{A}$, $\widehat{\boldsymbol\Theta}_1$ and $\widehat{\boldsymbol\Theta}_2$ until the stop criterion is satisfied, while the system rate is maximum.

In the following, the total computational complexity of Algorithm 3 is calculated as
\begin{align}
&\mathcal{O}\{ D_3( N^3+4M^3+4M^2N+2MN^2+2M^2K+8M^2+6N^2+9MN+5MK+5M+11N+\nonumber\\
&3+D_4(7N^3+27N^2+43N+18)) \}
\end{align}
FLOPs, where $D_3$ is the maximum number of alternating iterations for Algorithm 3 and $D_4$ is the number of iteration in GPI algorithm. Its highest order of computational complexity is $M^3$ and $N^3$ FLOPs, which is lower than the complexity of Algorithm 1.
\begin{table}[h]\footnotesize
\renewcommand{\arraystretch}{1}
\centering
\begin{tabular}{p {440pt}}
\hline
$\bf{Algorithm~3}$  Proposed WF-GPI-GRR Method\\
\hline
1.~~ Calculate $\beta_1$ and $\beta_2$ through (\ref{beta1}) and (\ref{beta2}).\\
2.~~ Initialize $\textbf{A}^0$, $\widehat{\boldsymbol\Theta}_1^0$ and $\widehat{\boldsymbol\Theta}_2^0$. According to (\ref{R}) and (\ref{SNR_3}), $R^0$ can be obtained.\\
3.~~ set the convergence error $\delta$ and the iteration number $t = 0$. \\
4.~\bf{repeat}  \\
5.~~ Fix $\widehat{\boldsymbol\Theta}_1^t$ and $\widehat{\boldsymbol\Theta}_2^t$, compute $\textbf{A}^{t+1}$ through (\ref{A1}).\\
6.~~ Fix $\textbf{A}^{t+1}$ and $\widehat{\boldsymbol\Theta}_2^t$, solve problem (\ref{v1_hat}) to achieve $\widehat{\textbf{v}}_1^{t+1}$ based on GPI presented in Algorithm 2, $\widehat{\boldsymbol\Theta}_1^{t+1}= \text{diag}\{ \widehat{\textbf{v}}_1^{t+1}(1:N)\}$.\\
7.~~ Fix $\textbf{A}^{t+1}$ and $\widehat{\boldsymbol\Theta}_1^{t+1}$, solve problem (\ref{v2_hat}) to achieve $\widehat{\textbf{v}}_2^{t+1}$ based on GRR theorem, $\widehat{\boldsymbol\Theta}_2^{t+1}= \text{diag}\{ \widehat{\textbf{v}}_2^{t+1}(1:N)\}$.\\
8.~~ Update $R^{t + 1}$ by using $\beta_1$, $\beta_2$, $\textbf{A}^{t + 1}$, $\widehat{\boldsymbol\Theta}_1^{t + 1}$ and $\widehat{\boldsymbol\Theta}_2^{t + 1}$.\\
9.~~ Update $t = t + 1$.\\
10.~\bf{until} \\
~~~~ $\left| {R^{t+1}- R^t} \right|\le \delta$.\\
\hline
\end{tabular}
\end{table}

\section{Simulation and Numerical Results}\label{Results}
\begin{figure*}[ht]
\begin{minipage}[t]{0.33\linewidth}
\centering
\includegraphics[width=2.25in, height=1.35in]{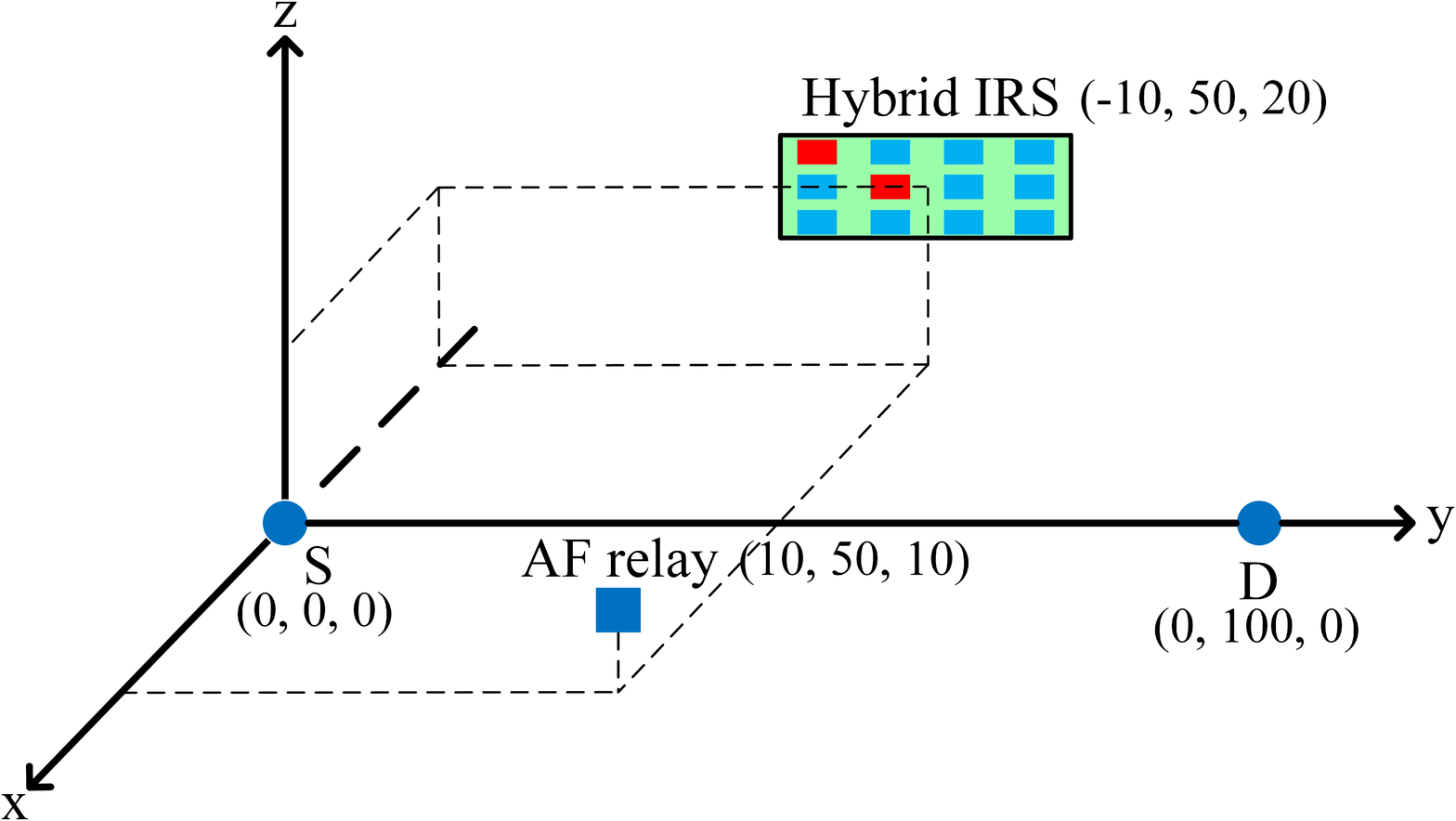}
\caption{Simulation setup.}\label{Simulation_setup}
\end{minipage}
\begin{minipage}[t]{0.33\linewidth}
\centering
\includegraphics[width=2.25in, height=1.90in]{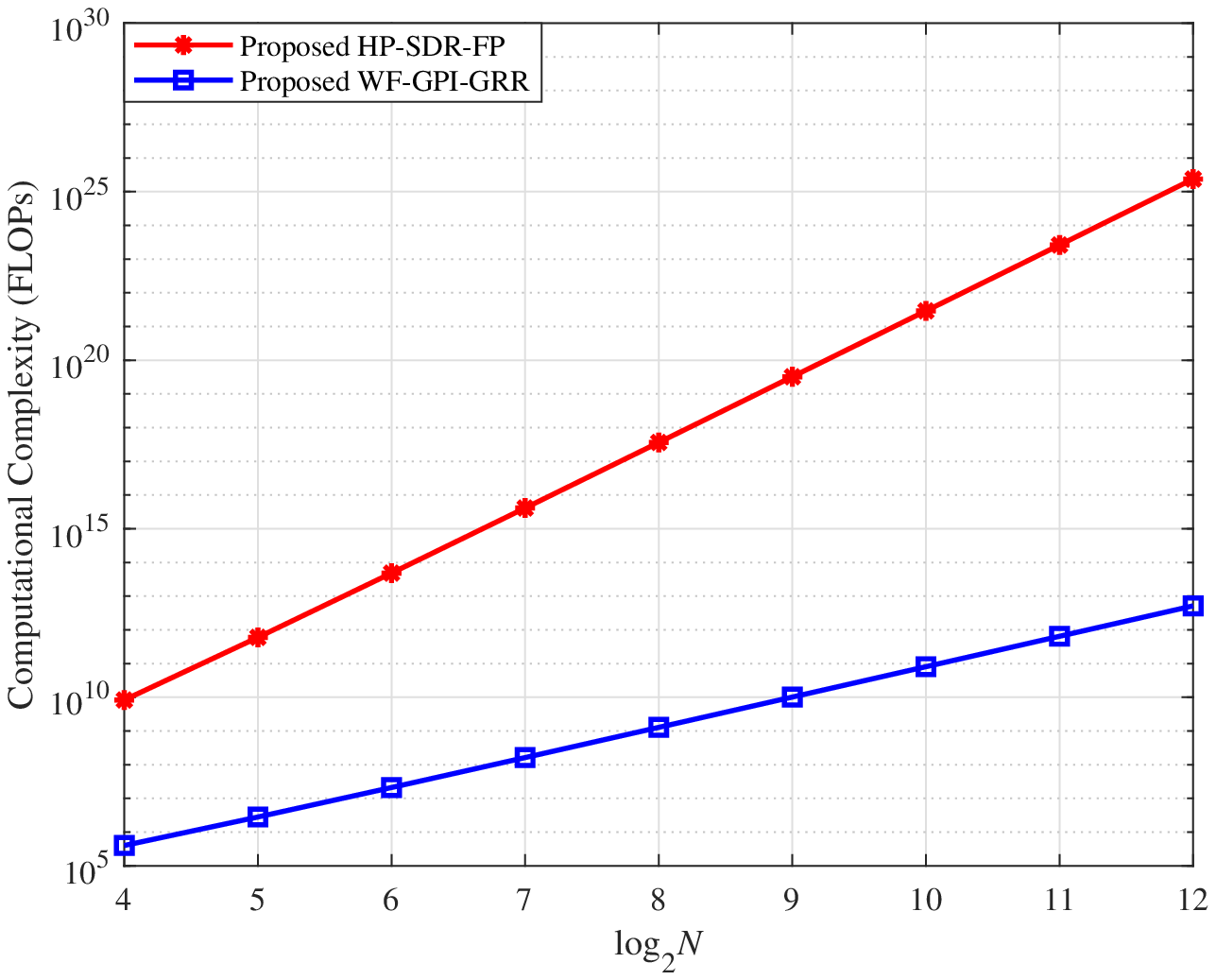}
\caption{Computational complexity versus $N$ with (M, K, D1, D2, D3, D4) = (2, 4, 6, 10, 5, 2).}\label{Complexity}
\end{minipage}
\begin{minipage}[t]{0.33\linewidth}
\centering
\includegraphics[width=2.25in, height=1.90in]{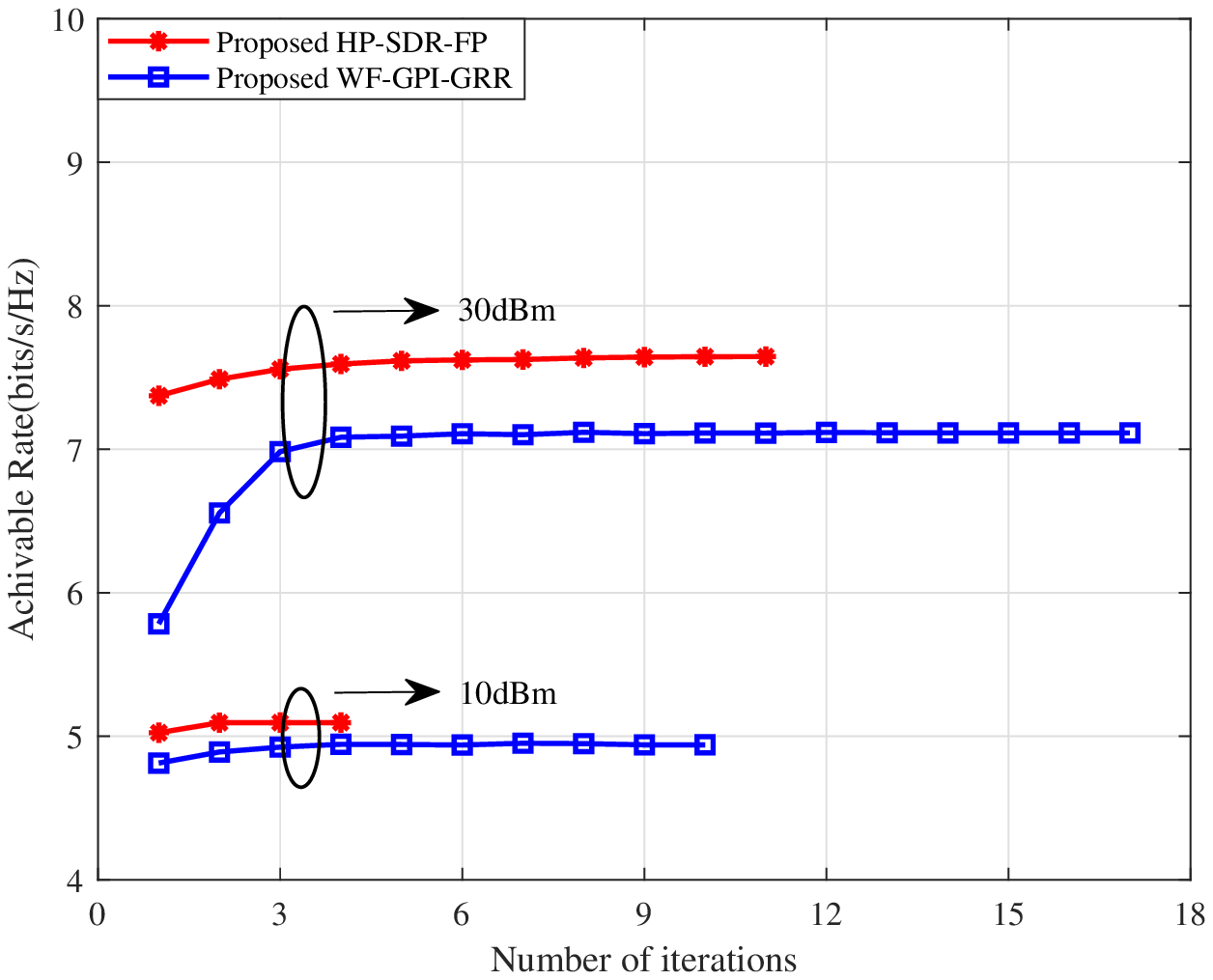}
\caption{Convergence of proposed methods with $(M, N, K, P_i, P_r)$ $=$ (2, 32, 4, 30dBm, 30dBm).}\label{iteration}
\end{minipage}
\end{figure*}

In this section, numerical simulations are performed to demonstrate the rate performance of the proposed HP-SDR-FP and WF-GPI-GRR methods.
As shown in Fig.~\ref{Simulation_setup}, the locations of S, D, hybrid IRS and AF relay are respectively set as (0, 0, 0), (0, 100m, 0), ($-$10m, 50m, 20m) and (10m, 50m, 10m). The path loss is modeled as $PL(d)=PL_0-10{\alpha}\text{log}_{10}(\frac{d}{d_0})$, where $PL_0$ is the path loss at the reference distance $d_0$. In generally, $PL_0= -$30dB and $d_0=1$m. $\alpha$ is the attenuation factor and $d$ is the distance between transceivers. In this paper, the attenuation factors of each channel associated with IRS, i.e., S-IRS, IRS-AF relay and IRS-D, are set as 2.0, and those of S-AF relay and AF relay-D links are considered as 3.0. The remaining system parameters are set as follow: $\sigma^2= -$80dBm and the sparse diagonal matrix $\textbf{E}_K$ is randomly generated.

Additionally, to demonstrate the proposed two methods, the following three benchmark schemes are taken into account.

1) ~$\textbf{AF relay+passive IRS}$: A passive IRS-aided AF relay network is considered, where IRS only reflects the signal without amplifying the reflected signal, and the reflecting coefficient of each IRS element is set as 1;

2) ~$\textbf{AF relay+passive IRS with random phase}$: With random phase of each reflection element uniformly and independently generated from the interval (0, $2\pi$], the beamforming matrix $\textbf{A}$ at AF relay is optimized.

3) $\textbf{Only AF relay}$: A AF relay network without IRS is considered, while the AF relay beamforming matrix $\textbf{A}$ can be achieved by MRC-MRT, which is given by $\textbf{A}=\sqrt{\frac{P_r}{P_s\|\boldsymbol\Gamma\textbf{h}_{sr}\|^2+\sigma^2\|\boldsymbol\Gamma\|_F^2}}\boldsymbol\Gamma$,
where $\boldsymbol\Gamma= \frac{\textbf{h}_{rd}\textbf{h}_{sr}^H}{\|\textbf{h}_{rd}^H\|\|\textbf{h}_{sr}\|}$.

Towards a fair comparison between a hybrid IRS-aided AF relay wireless network and the above three benchmark schemes, let us define that the total transmit power budgets of S and AF relay in the three benchmark schemes are the same as that of S, AF relay and IRS in the hybrid IRS-aided AF relay network. For instance, the AF relay transmit power budget $P_R$ in the three benchmark schemes is equal to $P_i + P_r$.

%\begin{figure}[h]
%\centering
%\includegraphics[width=0.450\textwidth,height=0.190\textheight]{Simulation_setup.eps}\\
%\caption{ Simulation setup.}\label{Simulation_setup.eps}
%\end{figure}
%
%\begin{figure}[h]
%\centering
%\includegraphics[width=0.450\textwidth,height=0.280\textheight]{Complexity.eps}\\
%\caption{  Computational complexity versus $N$ with (M, K, D1, D2, D3, D4) = (2, 4, 6, 10, 5, 2). }\label{Complexity.eps}
%\end{figure}

Fig.~\ref{Complexity} plots the computational complexity of the proposed two methods, namely HP-SDR-FP and WF-GPI-GRR. By suppressing $\text{ln}(1/\varepsilon)$ \cite{2019ZXB}, the computational complexities of the proposed two methods increase as $N$ increases. Clearly, the computational complexity of the HP-SDR-FP method is much higher than that of the WF-GPI-GRR method.

%\begin{figure}[h]
%\centering
%\includegraphics[width=0.470\textwidth,height=0.280\textheight]{iteration.eps}\\
%\caption{  Convergence of proposed methods with $(M, N, K, P_i, P_r)$ $=$ (2, 32, 4, 30dBm, 30dBm).}\label{iteration.eps}
%\end{figure}
%
%
%\begin{figure}[h]
%\centering
%\includegraphics[width=0.470\textwidth,height=0.280\textheight]{Rate_Vs_Ps.eps}\\
%\caption{  Achievable rate versus $P_s$ with $(M, N, K)$ $= (2, 32, 4)$.}\label{Rate_Vs_Ps.eps}
%\end{figure}

Fig.~\ref{iteration} demonstrates the proposed two methods are convergent under different $P_s$, respectively. Obviously, for $P_s=$ 10dBm, the proposed HP-SDR-FP and WF-GPI-GRR methods require about only four iterations to achieve the rate ceil. While for $P_s=$ 30dBm, it takes ten iterations for the proposed two methods to converge to the rate ceil. From the above two cases, we conclude that the proposed two methods are feasible.

\begin{figure*}[ht]
\begin{minipage}[t]{0.33\linewidth}
\centering
\includegraphics[width=2.25in, height=1.90in]{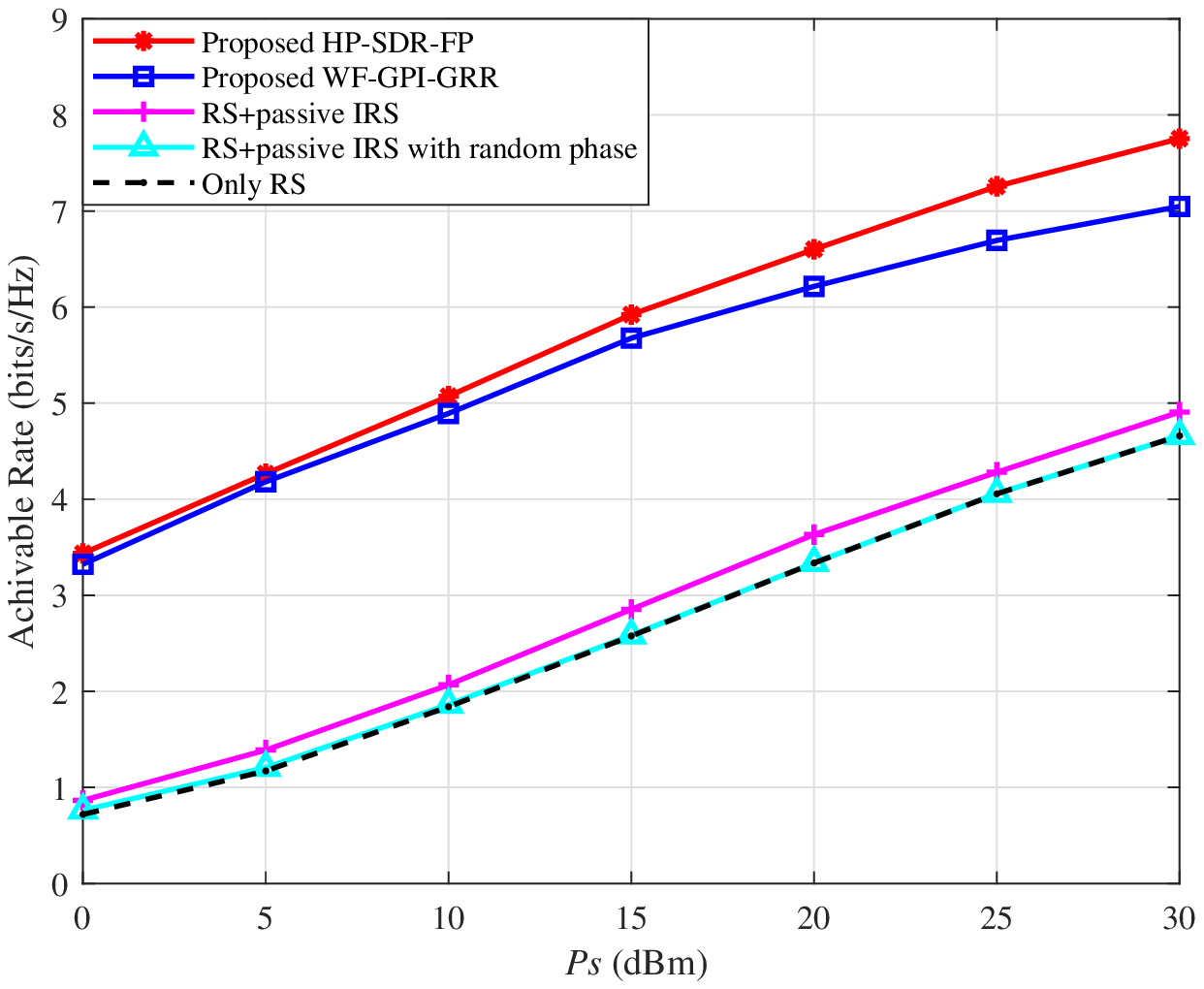}
\caption{Achievable rate versus $P_s$ with $(M, N, K)$ $= (2, 32, 4)$.}\label{Rate_Vs_Ps}
\end{minipage}
\begin{minipage}[t]{0.33\linewidth}
\centering
\includegraphics[width=2.25in, height=1.90in]{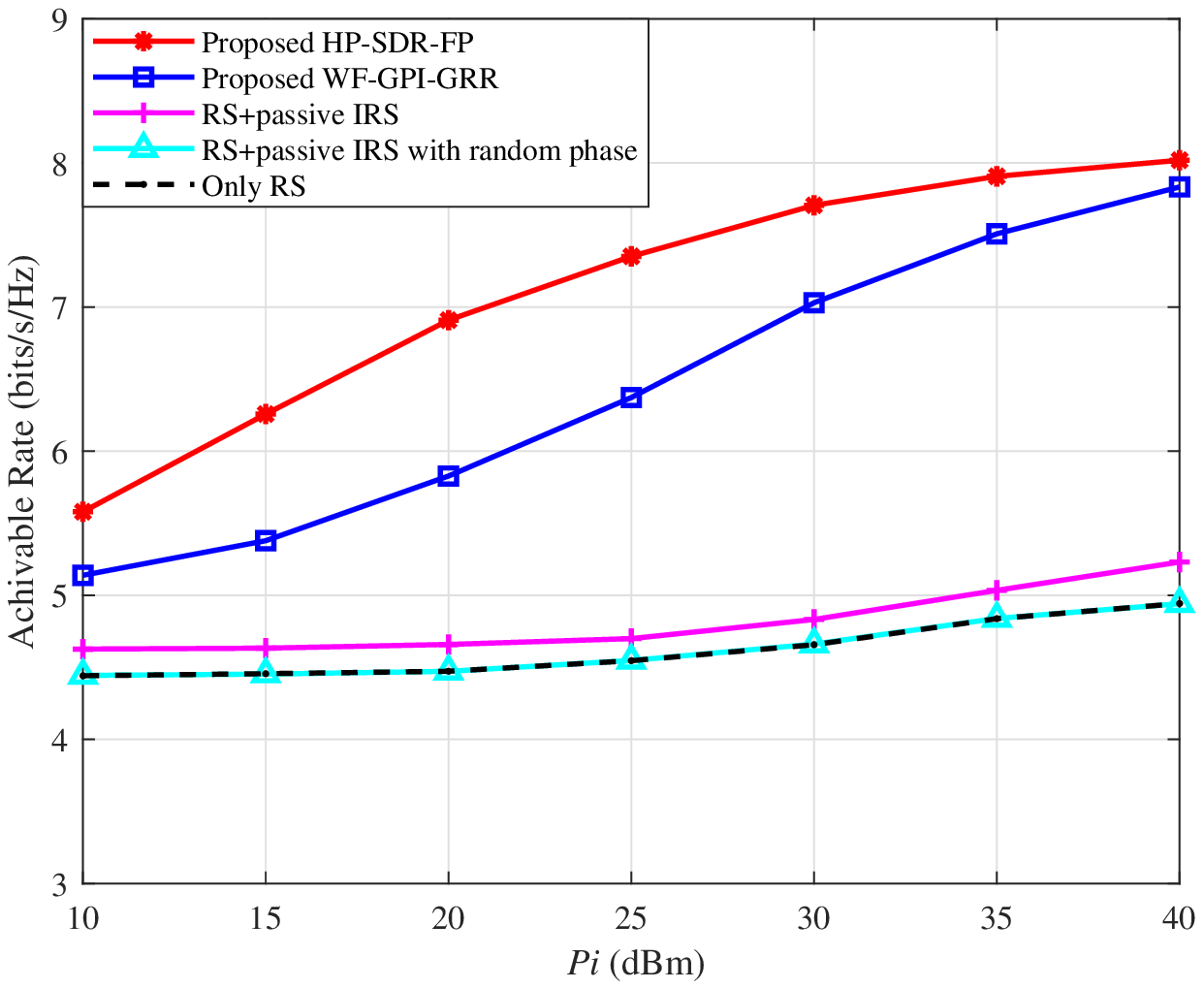}
\caption{Achievable rate versus $Pi$ with $(M, N, K, P_s)$ $=$ (2, 32, 4, 30dBm).}\label{Rate_VS_Pi}
\end{minipage}
\begin{minipage}[t]{0.33\linewidth}
\centering
\includegraphics[width=2.25in, height=1.90in]{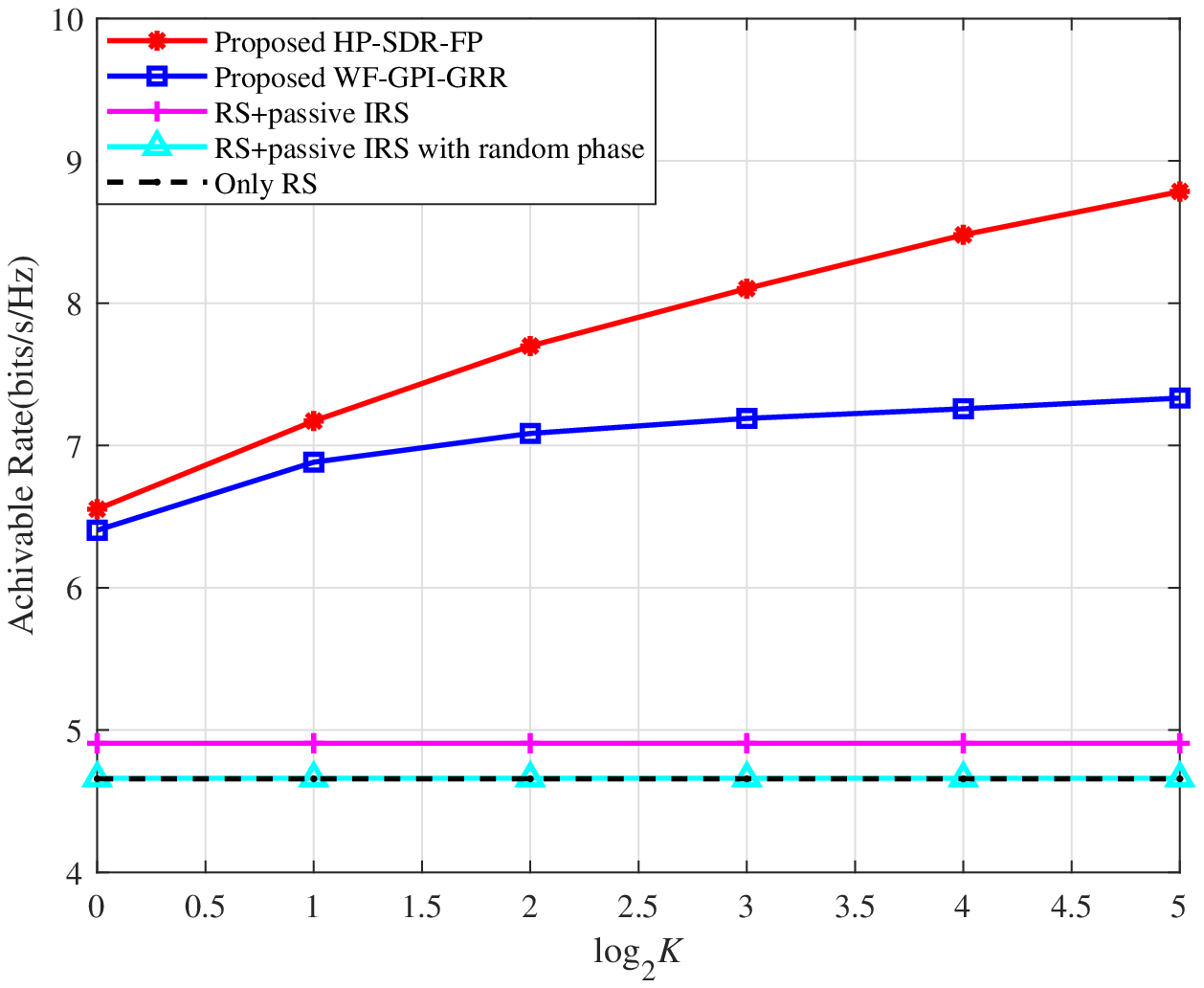}
\caption{Achievable rate versus $K$ with $(M, N, P_s)$ $=$ (2, 32, 30dBm).}\label{Rate_Vs_K}
\end{minipage}
\end{figure*}

Fig.~\ref{Rate_Vs_Ps} plots the curves of achievable rate versus $P_s$. It is clear that regardless of the proposed HP-SDR-FP and WF-GPI-GRR methods or the three benchmarks, their rates can be improved along with the increase of $P_s$. Meanwhile, the rates obtained by the proposed two methods with $(P_i, P_r)$ $=$ (30dBm, 30dBm) surpass those of the three benchmarks with $P_R$ $=$ 33dBm in all $P_s$ region.
Furthermore, the rate performance of WF-GPI-GRR method is close to that of HP-SDR-FP method in the low power $P_s$ region. Compared to the three benchmarks, the proposed two methods can respectively obtain 145.26\% and 136.61\% rate gains when $P_s$ $=$ 10dBm.

%\begin{figure}[h]
%\centering
%\includegraphics[width=0.470\textwidth,height=0.280\textheight]{Rate_VS_Pi.eps}\\
%\caption{  Achievable rate versus $Pi$ with $(M, N, K, P_s)$ $=$ (2, 32, 4, 30dBm).}\label{Rate_VS_Pi.eps}
%\end{figure}
%
%
%\begin{figure}[h]
%\centering
%\includegraphics[width=0.470\textwidth,height=0.280\textheight]{Rate_Vs_K.eps}\\
%\caption{  Achievable rate versus $K$ with $(M, N, P_s)$ $=$ (2, 32, 30dBm).}\label{Rate_Vs_K.eps}
%\end{figure}

Fig.~\ref{Rate_VS_Pi} shows the curves of achievable rate versus $P_i$. With the growth of $P_i$, it is observed that the proposed HP-SDR-FP and WF-GPI-GRR methods can achieve significant rate improvement than those of the three benchmarks in the case of $(P_r, P_R)$ $=$ (30dBm, 33dBm). Particularly, when $P_i$ is equal to 40dBm, the proposed HP-SDR-FP and WF-GPI-GRR methods can respectively harvest up to 53.3\% and 49.8\% rate gains over AF relay+passive IRS.

Fig.~\ref{Rate_Vs_K} presents the curves of achievable rate versus the number $K$ of active IRS elements. Obviously, as the number of active IRS elements $K$ increases, the rate gains obtained by the proposed two methods with $(P_i, P_r)$ $=$ (30dBm, 30dBm) increase gradually and become more significant over the three benchmark schemes with $P_R$ $=$ 33dBm. Compared with the benchmark scheme of AF relay+passive IRS, our proposed two methods perform much better, which shows that the optimization of beamforming is important and efficient.

\section{Conclusions}\label{Conclusions}
In this paper, a hybrid IRS-aided AF relay network was investigated, where the hybrid IRS was made up of active units and passive units. To achieve a goal of rate performance enhancement, two efficient beamforming methods called HP-SDR-FP and WF-GPI-GRR were proposed to maximize SNR by alternately designing the beamforming matrix and reflecting coefficient matrices of two time slots. Simulation results show that the proposed two methods can obtain appreciable rate gain over the passive IRS-assisted AF relay network and the AF relay network. It verifies the active IRS elements can break the ``double fading'' effect caused by conventional passive IRS. For instance, an approximate 50.0\% rate gain over the three benchmark schemes can be achieved in the high power budget region of hybrid IRS. Therefore, a hybrid IRS-aided AF relay network can provide an enhancement in accordance with rate performance and extended coverage for the mobile communications.

\ifCLASSOPTIONcaptionsoff
  \newpage
\fi

\bibliographystyle{IEEEtran}
\bibliography{IEEEfull,reference}

% Generated by IEEEtran.bst, version: 1.14 (2015/08/26)
\begin{thebibliography}{10}
\providecommand{\url}[1]{#1}
\csname url@samestyle\endcsname
\providecommand{\newblock}{\relax}
\providecommand{\bibinfo}[2]{#2}
\providecommand{\BIBentrySTDinterwordspacing}{\spaceskip=0pt\relax}
\providecommand{\BIBentryALTinterwordstretchfactor}{4}
\providecommand{\BIBentryALTinterwordspacing}{\spaceskip=\fontdimen2\font plus
\BIBentryALTinterwordstretchfactor\fontdimen3\font minus
  \fontdimen4\font\relax}
\providecommand{\BIBforeignlanguage}[2]{{%
\expandafter\ifx\csname l@#1\endcsname\relax
\typeout{** WARNING: IEEEtran.bst: No hyphenation pattern has been}%
\typeout{** loaded for the language `#1'. Using the pattern for}%
\typeout{** the default language instead.}%
\else
\language=\csname l@#1\endcsname
\fi
#2}}
\providecommand{\BIBdecl}{\relax}
\BIBdecl

\bibitem{2020LC}
L.~Chettri and R.~Bera, ``A comprehensive survey on internet of things {(IoT)}
  toward {5G} wireless systems,'' \emph{IEEE Internet Things J.}, vol.~7,
  no.~1, pp. 16--32, Jan. 2020.

\bibitem{2021CJN}
J.~Chen, S.~Li, J.~Xing, J.~Wang, and S.~Fu, ``Multiple nodes access of
  wireless beam modulation for 6g-enabled internet of things,'' \emph{IEEE
  Internet Things J.}, vol.~8, no.~20, pp. 15\,191--15\,204, Oct. 2021.

\bibitem{2017VWSW}
V.~W.~S. Wong, R.~Schober, D.~W.~K. Ng, and L.-C. Wang, \emph{Key Technologies
  for {5G} Wireless Systems}, Cambridge U.K.: Cambridge Univ. Press, 2017.

\bibitem{2017PS}
S.~Poursajadi, M.~H. Madani, and H.~K. Bizaki, ``Power allocation and outage
  probability analysis of {AF} relaying systems with multiple antennas at
  terminal nodes,'' \emph{IEEE Trans. Veh. Technol.}, vol.~66, no.~1, pp.
  377--384, Jan. 2017.

\bibitem{2020WQQ}
Q.~Wu and R.~Zhang, ``Beamforming optimization for wireless network aided by
  intelligent reflecting surface with discrete phase shifts,'' \emph{IEEE
  Trans. Commun.}, vol.~68, no.~3, pp. 1838--1851, Mar. 2020.

\bibitem{2021LY}
Y.~Liang, J.~Chen, R.~Long, Z.~He, X.~Lin, C.~Huang, S.~Liu, X.~Shen, and M.~D.
  Renzo, ``Reconfigurable intelligent surfaces for smart wireless environments:
  channel estimation, system design and applications in 6{G} networks,''
  \emph{Sci. China Inf. Sci.}, vol.~64, no.~10, pp. 1--21, Jul. 2021.

\bibitem{2021TWK}
W.~Tang, M.~Z. Chen, X.~Chen, J.~Y. Dai, Y.~Han, M.~Di~Renzo, Y.~Zeng, S.~Jin,
  Q.~Cheng, and T.~J. Cui, ``Wireless communications with reconfigurable
  intelligent surface: Path loss modeling and experimental measurement,''
  \emph{IEEE Trans. Wireless Commun.}, vol.~20, no.~1, pp. 421--439, Jan. 2021.

\bibitem{2020YL1}
L.~Yang, X.~Yan, D.~B. da~Costa, T.~A. Tsiftsis, H.-C. Yang, and M.-S. Alouini,
  ``Indoor mixed dual-hop {VLC/RF} systems through reconfigurable intelligent
  surfaces,'' \emph{IEEE Wireless Commun. Lett.}, vol.~9, no.~11, pp.
  1995--1999, Nov. 2020.

\bibitem{2022JXY}
F.~Shu, L.~Yang, X.~Jiang, W.~Cai, W.~Shi, M.~Huang, J.~Wang, and X.~You,
  ``Beamforming and transmit power design for intelligent reconfigurable
  surface-aided secure spatial modulation,'' \emph{IEEE J. Sel. Topics Signal
  Process.}, vol.~16, no.~5, pp. 933--949, Aug. 2022.

\bibitem{2021TY}
F.~Shu, Y.~Teng, J.~Li, M.~Huang, W.~Shi, J.~Li, Y.~Wu, and J.~Wang, ``Enhanced
  secrecy rate maximization for directional modulation networks via {IRS},''
  \emph{IEEE Trans. Commun.}, vol.~69, no.~12, pp. 8388--8401, Dec. 2021.

\bibitem{2019SH}
H.~Shen, W.~Xu, S.~Gong, Z.~He, and C.~Zhao, ``Secrecy rate maximization for
  intelligent reflecting surface assisted multi-antenna communications,''
  \emph{IEEE Commun. Lett.}, vol.~23, no.~9, pp. 1488--1492, Sep. 2019.

\bibitem{2020WQQ1}
Q.~Wu and R.~Zhang, ``Weighted sum power maximization for intelligent
  reflecting surface aided {SWIPT},'' \emph{IEEE Wireless Commun. Lett.},
  vol.~9, no.~5, pp. 586--590, May 2020.

\bibitem{2021SWP}
W.~Shi, X.~Zhou, L.~Jia, Y.~Wu, F.~Shu, and J.~Wang, ``Enhanced secure wireless
  information and power transfer via intelligent reflecting surface,''
  \emph{IEEE Commun. Lett.}, vol.~25, no.~4, pp. 1084--1088, Apr. 2021.

\bibitem{2020PCH}
C.~Pan, H.~Ren, K.~Wang, W.~Xu, M.~Elkashlan, A.~Nallanathan, and L.~Hanzo,
  ``Multicell {MIMO} communications relying on intelligent reflecting
  surfaces,'' \emph{IEEE Trans. Wireless Commun.}, vol.~19, no.~8, pp.
  5218--5233, Aug. 2020.

\bibitem{2022RA}
A.~Rezaei, A.~Khalili, J.~Jalali, H.~Shafiei, and Q.~Wu, ``Energy-efficient
  resource allocation and antenna selection for {IRS}-assisted multicell
  downlink networks,'' \emph{IEEE Wireless Commun. Lett.}, vol.~11, no.~6, pp.
  1229--1233, June 2022.

\bibitem{2022ZXB}
X.~Zhou, S.~Yan, Q.~Wu, F.~Shu, and D.~Ng, ``Intelligent reflecting surface
  ({IRS})-aided covert wireless communications with delay constraint,''
  \emph{IEEE Trans. Wireless Commun.}, vol.~21, no.~1, pp. 532--547, Jan. 2022.

\bibitem{2022CX}
X.~Chen, T.-X. Zheng, L.~Dong, M.~Lin, and J.~Yuan, ``Enhancing {MIMO} covert
  communications via intelligent reflecting surface,'' \emph{IEEE Wireless
  Commun. Lett.}, vol.~11, no.~1, pp. 33--37, Jan. 2022.

\bibitem{2022WQQ}
Q.~Wu, X.~Zhou, W.~Chen, J.~Li, and X.~Zhang, ``{IRS}-aided {WPCNs}: A new
  optimization framework for dynamic {IRS} beamforming,'' \emph{IEEE Trans.
  Wireless Commun.}, vol.~21, no.~7, pp. 4725--4739, July 2022.

\bibitem{2021CHQ}
H.~Cao, Z.~Li, and W.~Chen, ``Resource allocation for {IRS}-assisted wireless
  powered communication networks,'' \emph{IEEE Wireless Commun. Lett.},
  vol.~10, no.~11, pp. 2450--2454, Nov. 2021.

\bibitem{2022SWP}
W.~Shi, Q.~Wu, F.~Xiao, F.~Shu, and J.~Wang, ``Secrecy throughput maximization
  for {IRS}-aided {MIMO} wireless powered communication networks,'' \emph{IEEE
  Trans. Commun.}, vol.~70, no.~11, pp. 7520--7535, Nov. 2022.

\bibitem{2020YL2}
L.~Yang, W.~Guo, and S.~I. Ansari, ``Mixed dual-hop fso-rf communication
  systems through reconfigurable intelligent surface,'' \emph{IEEE Commun.
  Lett.}, vol.~24, no.~7, pp. 1558--1562, July 2020.

\bibitem{2021YIK}
I.~Yildirim, F.~Kilinc, E.~Basar, and G.~C. Alexandropoulos, ``Hybrid
  {RIS}-empowered reflection and decode-and-forward relaying for coverage
  extension,'' \emph{IEEE Commun. Lett.}, vol.~25, no.~5, pp. 1692--1696, May
  2021.

\bibitem{2021MO}
M.~Obeed and A.~Chaaban, ``Relay-reconfigurable intelligent surface cooperation
  for energy-efficient multiuser systems,'' in \emph{2021 IEEE Int. Conf.
  Commun. Workshops (ICC Workshops)}, pp. 1--6.

\bibitem{2021NTN}
N.~T. Nguyen, Q.-D. Vu, K.~Lee, and M.~Juntti, ``Spectral efficiency
  optimization for hybrid relay-reflecting intelligent surface,'' in \emph{2021
  IEEE Int. Conf. Commun. Workshops (ICC Workshops)}, Jul. 2021, pp. 1--6.

\bibitem{2022WXH}
X.~Wang, F.~Shu, W.~Shi, X.~Liang, R.~Dong, J.~Li, and J.~Wang, ``Beamforming
  design for {IRS}-aided decode-and-forward relay wireless network,''
  \emph{IEEE Trans. Green Commun. Netw.}, vol.~6, no.~1, pp. 198--207, Mar.
  2022.

\bibitem{2020AZD}
Z.~Abdullah, G.~Chen, S.~Lambotharan, and J.~A. Chambers, ``A hybrid relay and
  intelligent reflecting surface network and its ergodic performance
  analysis,'' \emph{IEEE Wireless Commun. Lett.}, vol.~9, no.~10, pp.
  1653--1657, Oct. 2020.

\bibitem{2022ZZJ}
Z.~Zhang, L.~Dai, X.~Chen, C.~Liu, F.~Yang, R.~Schober, and H.~Vincent~Poor,
  ``Active {RIS} vs. passive {RIS}: Which will prevail in {6G}?'' \emph{IEEE
  Trans. Commun.}, vol.~71, no.~3, pp. 1707--1725, 2023.

\bibitem{2022ZKD}
K.~Zhi, C.~Pan, H.~Ren, K.~K. Chai, and M.~Elkashlan, ``Active {RIS} versus
  passive {RIS}: Which is superior with the same power budget?'' \emph{IEEE
  Commun. Lett.}, vol.~26, no.~5, pp. 1150--1154, May 2022.

\bibitem{2022DLM}
L.~Dong, H.-M. Wang, and J.~Bai, ``Active reconfigurable intelligent surface
  aided secure transmission,'' \emph{IEEE Trans. Veh. Technol.}, vol.~71,
  no.~2, pp. 2181--2186, Feb. 2022.

\bibitem{2022MN}
N.~Mensi and D.~B. Rawat, ``Reconfigurable intelligent surface selection for
  wireless vehicular communications,'' \emph{IEEE Wireless Commun. Lett.},
  vol.~11, no.~8, pp. 1743--1747, Aug. 2022.

\bibitem{2019ZXB}
X.~Zhou, J.~Li, F.~Shu, Q.~Wu, Y.~Wu, W.~Chen, and L.~Hanzo, ``Secure {SWIPT}
  for directional modulation-aided {AF} relaying networks,'' \emph{IEEE J. Sel.
  Areas Commun.}, vol.~37, no.~2, pp. 253--268, Feb. 2019.

\end{thebibliography}
\end{document}